\documentclass[11pt]{article}
\usepackage[DIV12]{typearea}
\usepackage{color}
\usepackage{amsmath,amssymb}
\usepackage{amsfonts}  
\usepackage{mathrsfs}
\usepackage{graphicx} 
\usepackage{bbm}
\usepackage{longtable}
\usepackage[small]{caption2} 
\usepackage{fleqn} 
\usepackage{cite}
\usepackage{pifont}
\usepackage{enumerate} 
\usepackage{pdflscape}
\usepackage{hhline}  
\usepackage{multirow} %
\usepackage{booktabs} 
\usepackage{tabulary}

\newcommand{\I}{\mathrm{i}}

\newcommand{\E}[1]{\ensuremath{\mathrm{E}_{#1}}} 

\newcommand{\OO}[1]{\ensuremath{\mathrm{O}(#1)}}
\newcommand{\SO}[1]{\ensuremath{\mathrm{SO}(#1)}}
\newcommand{\SU}[1]{\ensuremath{\mathrm{SU}(#1)}}
\newcommand{\U}[1]{\ensuremath{\mathrm{U}(#1)}}
\newcommand{\Z}[1]{\ensuremath{\mathbbm{Z}_{#1}}} 
\newcommand{\ZxZ}[2]{\ensuremath{\mathbbm{Z}_{#1}\times\mathbbm{Z}_{#2}}} 
\newcommand{\QQ}{\ensuremath{\mathbbm{Q}}}  

\newcommand{\bs}[1]{\ensuremath{\boldsymbol{#1}}}

\newcommand{\maG}{\ensuremath{\mathcal{G}} }
\newcommand{\maN}{\ensuremath{\mathcal{N}} }

\newcommand{\x}{\ensuremath{\times}}
\newcommand{\id}{\ensuremath{\mathbbm{1}}}

\newcolumntype{R}[1]{>{\raggedleft\arraybackslash}p{#1}} 

\allowdisplaybreaks

\begin{document}

\date{}
\title{
  \vskip 2cm
  {\bf\huge Charting the flavor landscape\\ of MSSM-like Abelian heterotic orbifolds}\\[0.8cm]
}
\author{
 {\bf\normalsize
  Yessenia Olgu\'in--Trejo$^1$\footnote{\texttt{yess.olt@ciencias.unam.mx}},
  Ricardo P\'erez--Mart\'inez$^{1,2}$\footnote{\texttt{ricardoperezm@estudiantes.fisica.unam.mx}} and
  Sa\'ul~Ramos--S\'anchez$^1$\footnote{\texttt{ramos@fisica.unam.mx}}
 }\\[5mm]
 {\it\normalsize $^1$Instituto de F\'isica, Universidad Nacional Aut\'onoma de M\'exico,}\\
 {\it\normalsize POB 20-364, Cd.Mx. 01000, M\'exico}\\[5mm]
 {\it\normalsize $^2$Facultad de Ciencias F\'isico-Matem\'aticas, Universidad Aut\'onoma de Coahuila,}\\
 {\it\normalsize Edificio A, Unidad Camporredondo, 25000, Saltillo, Coahuila, M\'exico}
}

\maketitle 

\thispagestyle{empty}

\vskip 1cm
\begin{abstract}
Discovering a selection principle and the origin of flavor symmetries from an ultraviolet completion of particle physics
is an interesting open task. As a step in this direction, we classify all possible flavor symmetries 
of 4D massless spectra emerging from supersymmetric Abelian orbifold compactifications, including roto-translations
and non-factorizable tori, for generic moduli values. Although these symmetries are valid in 
all string theories, we focus on the \E8\x\E8 heterotic string. 
We perform the widest known search of \E8\x\E8 Abelian orbifold compactifications, yielding over 121,000 models 
with MSSM-like features. About 75.4\% of these models exhibit flavor symmetries containing $D_4$ factors        
and only nearly 1.2\% have $\Delta(54)$ factors.                                                                
The remaining models are furnished with purely Abelian flavor symmetries. 
Our findings suggest that, should particle phenomenology arise from such a heterotic orbifold,
it could accommodate only one of these flavor symmetries.
\end{abstract}

\clearpage
\newpage

\section{Introduction}

The reason of the number of families in the standard model (SM) as well as the origin of fermion mixings
may be clarified in extensions of the SM.
The general structure of the quark-mixing matrix motivated the bottom-up introduction of {\it ad hoc} 
discrete flavor symmetries (see e.g.~\cite{Ishimori:2010au,King:2013eh} for a review) that, together with 
a number of extra fields transforming in non-trivial flavor representations, yield new phenomenology 
that may be contrasted with observations. Choosing the correct flavor symmetry among the different
scenarios that render similar physics requires a selection principle that is not found in this
field-theoretic approach.

It is in this sense that, given the constraints of string theory and its potential to provide
an ultraviolet completion of the SM, we can try to identify a mechanism  in string theory to 
restrict the admissible flavor symmetries, providing thereby their origin. This quest is not new.
The seminal works were in the context of heterotic orbifold compactifications~\cite{Kobayashi:2006wq,Nilles:2012cy},
which sparked the study of the phenomenological consequences of some 
models~\cite{Kobayashi:2004ud,Ko:2007dz,Pena:2012ki,Carballo-Perez:2016ooy,Kim:2018ynz}, generalizations in models with 
magnetic fluxes~\cite{Abe:2016eyh} and relations with modular symmetries~\cite{Kobayashi:2018rad}. 
Flavor symmetries are associated in these works with geometric aspects of orbifolds, but
they can also be related to larger continuous symmetries of the extra dimensions~\cite{Beye:2014nxa}.
Also in D-brane compactifications, some sources of flavor symmetries have been identified and there is
progress in the study of their phenomenology~\cite{Abe:2009vi,BerasaluceGonzalez:2012vb,Marchesano:2013ega,Hamada:2014hpa}.

Here we focus on the \E8\x\E8 heterotic string compactified on {\it all} symmetric, toroidal, 
Abelian orbifolds\footnote{Beside the original works~\cite{Dixon:1985jw,Dixon:1986jc}, there are several 
good introductions to these constructions, see e.g.~\cite{Bailin:1999nk,RamosSanchez:2008tn,Vaudrevange:2008sm}.} 
that yield 4D $\maN=1$ low-energy effective field theories, recently 
classified in ref.~\cite{Fischer:2012qj}. In these scenarios, the fact that most matter states
are localized at the curvature singularities of the orbifold becomes instrumental to arrive at
flavor phenomenology, because different singularities are assigned different localization numbers
that can be interpreted as charges of a flavor symmetry in the 4D resulting model. 

In this work, we present first a systematic classification of flavor symmetries in Abelian toroidal orbifolds, 
whose moduli have no special values, avoiding possible enhancements. These symmetries are completely determined 
by the orbifold space group, whose nature is purely geometric, and are thus independent of the 
string theory to be compactified. As the geometric structure of a toroidal orbifold can be more complicated 
than usually assumed, due to the presence of roto-translations or non-factorizable tori, this task can 
be challenging and lead to flavor symmetries not yet identified.
Since we explore here all 6D orbifolds, this paper represents the completion of
the work initiated in ref.~\cite{Ramos-Sanchez:2017lmj}.

Orbifolds are used in the heterotic strings to obtain 
models that reproduce the main features of the SM~\cite{Blaszczyk:2014qoa}, 
its minimal supersymmetric extension~\cite{Kobayashi:2004ya,Buchmuller:2005jr,Kim:2007mt,Blaszczyk:2009in} (MSSM)
and other non--minimal extensions~\cite{Lebedev:2009ag}, as well as many other observed and/or desirable properties
of particle physics~\cite{Buchmuller:2007zd,Kappl:2008ie,Choi:2009jt,Brummer:2010fr,Krippendorf:2012ir,Badziak:2012yg,Kim:2015mpa}.
Aiming at gaining insight on the actual flavor symmetry of Nature, an interesting question we can pursue is: 
what flavor symmetries can these orbifolds have?

To answer this question, we perform a search of semi-realistic $\maN=1$  heterotic orbifolds
with help of the \texttt{orbifolder}~\cite{Nilles:2011aj}. We then study their flavor symmetries, which
build subgroups of the symmetries we classify in section~\ref{sec:symmetries}.
We expect that a statistical analysis of these findings may hint
towards the family structure that particle physics emerging from strings can have.


This paper is organized as follows. After reviewing the aspects of heterotic orbifolds
that are crucial for our study on flavor symmetries, we proceed in section~\ref{sec:symmetries}
to discuss how flavor symmetries arise in Abelian toroidal orbifolds. We then classify all 
flavor symmetries that can arise from these orbifolds, independently of the string theory
one may compactify. In section~\ref{sec:class}, we show the results of the most comprehensive
search of semi-realistic heterotic orbifolds so far. Section~\ref{sec:stringyFlavors} is devoted
to the discussion of the flavor symmetries that arise in the promising models we found, which are 
summarized in the tables presented in the appendix. In section~\ref{sec:conclusions}, we provide our
summary and outlook.


\section{Orbifold compactifications}
\label{sec:orbifolds}

\subsection{Toroidal orbifolds}

In order to introduce our notation and the main aspects of our constructions,
let us first study the structure of 6D toroidal orbifolds in the context of 4D $\maN=1$
models resulting from the supersymmetric heterotic strings.

In general, a 6D toroidal orbifold $\mathcal{O}$ is defined as the quotient space that results from dividing a 6D
torus $\mathbb{T}^{6}$ by the so-called orbifolding group $G$. The torus can be embedded in $\mathbb{R}^6$ by dividing this space
by a lattice $\Lambda$ with basis vectors $\{e_i | i=1,\ldots,6\}$, corresponding to identifying all points of
$\mathbb{R}^6$ connected by translations $\lambda\in\Lambda$, such that $\lambda = \sum_i m_i e_i$ for some integers $m_i$.

Alternatively, one can produce the same orbifold $\mathcal{O}$ by moding $\mathbb{R}^6$ 
by the space group $S$, which is a discrete group of isometries of the torus $\mathbb{T}^6$, including
the translations in the lattice $\Lambda$. For our
purposes, this description of an orbifold turns out to be more useful. That is, we shall consider here a 
6D toroidal orbifold defined as
\begin{equation}
  \mathcal{O} = \mathbb{R}^6 / S\,.
\end{equation}
The elements $g\in S$ have the general structure $g=(\vartheta,\mu)$, where the operators
$\vartheta$ are in general elements of $\OO{6}$ that form a discrete, Abelian or non-Abelian 
point group $P$ of $S$,
and $\mu$ is a vector in $\mathbb{R}^6$, which may or may not be an element of the torus lattice,
although it can always be written in the basis of $\Lambda$ (with arbitrary coefficients). 
The action of $g\in S$ on $x\in\mathbb{R}^6$ is defined by
\begin{equation}
\label{eq:gonX}
  x \ \stackrel{g}{\longmapsto} \ g x = \vartheta x + \mu\,,
\end{equation}
that is, $\vartheta$ denotes a rotation, reflection or inversion of $x$ whereas $\mu$ denotes a
translation vector. 

It is said that the action of $g$ is trivial on the torus only if it amounts to
a lattice translation. This is because $\mathbb{T}^{6} = \mathbb{R}^6 / \Lambda$, i.e. the torus
is obtained by the identification $x\simeq x+\lambda$, $\lambda\in\Lambda$.
It follows that, if $\mu$ is an element of the torus lattice, $\mu=\lambda\in\Lambda$, 
the only component of $g\in S$ that exerts a non-trivial action on the torus is $\vartheta$,
since $\vartheta x$ and $\vartheta x + \lambda$ are identified on a {\it toroidal} orbifold. 

When $\mu$ in eq.~\eqref{eq:gonX} is chosen to be a more general vector, $\mu\notin\Lambda$, 
the space-group element $g=(\vartheta,\mu)$ is called a {\it roto-translation}. In this case,
both $\vartheta$ and $\mu$ act non-trivially on the 6D torus. One of the purposes of this work
is to study this case with more attention, attempting to pave the path towards phenomenology of
orbifolds with roto-translations.

In an orbifold, the space group defining the orbifold consists of a finite number of elements $g\in S$ 
called space group {\it generators}, their products, computed according to
\begin{equation}
\label{eq:Sproduct}
 g'' = g g' = (\vartheta,\mu)(\vartheta',\mu') = (\vartheta\vartheta',\mu + \vartheta\mu'),\qquad g,g',g''\in S\,,
\end{equation}
and their conjugations. All elements of a space group $S$ 
can be grouped in different conjugacy classes $[g]=\{h^{-1}gh, h\in S\}$. 
All elements of a conjugacy class are equivalent. 
Note that an element $g=(\vartheta,\mu + \sum_i m_i e_i)$, with $\mu\notin\Lambda$ or null, 
can be rewritten as $\prod_i(\id,e_i)^{m_i} (\vartheta,\mu)$.
Therefore, the space group generators can be pure \OO6 transformations, roto-translations
or translations.

An additional property of orbifold generators is that each of them has an integer order $N$, such that
$g^N$ is trivial on the torus, that is $g^N = (\id, \lambda)$ with $\lambda\in\Lambda$.
We point out that this restricts the shape of the translation vectors $\mu\notin\Lambda$ 
of roto-translations $g=(\vartheta,\mu)$. The trivial action of $g^N$ on the torus
implies that $\vartheta^N=\id$ and $\sum_{j=0}^{N-1} \vartheta^j \mu\in\Lambda$.
Notice that, for example, if $\vartheta\mu=\mu\neq0$, then the translation vector 
is given as a fraction of lattice vector, $\mu=\frac{1}{N}\lambda$.

Let us focus now on Abelian orbifolds, which are the scope of this work. 
Complexifying the orbifold generators $g$, eq.~\eqref{eq:gonX} becomes
\begin{equation}
\label{eq:gonZ}
  z \ \stackrel{g}{\longmapsto} \ g z = \vartheta z + \mu\,,\qquad z,\mu\in\mathbb{C}^3\,,
\end{equation}
with the complex coordinates of $z$ related by $z^a=x^{2a-1}+ix^{2a}$, $a=1,2,3$, with the real
coordinates $x\in\mathbb{R}^6$. In Abelian orbifolds, the complexified $\vartheta$ elements of the 
space group generators can be simultaneously diagonalized and represented as matrices of the form 
$\vartheta=\text{diag}(e^{2\pi\I v_1},e^{2\pi\I v_2}, e^{2\pi\I v_3})$, with $0\leq|v_a|<1$.
The vector $v=(v_1,v_2,v_3)$ is commonly called twist vector.

\subsubsection{Fixed points and roto-translations}
\label{sec:fixedpoints}

Space group generators $g$ with non-trivial twist $\vartheta$ have a non-free action on $\mathbb T^6$. This implies that,
in these cases, some points are left unaltered or fixed under $g$, which correspond to curvature singularities
of the compact space. The simplest example of such a fixed point is $z=0$ for the space group element 
$g=(\vartheta,0)$ with $\vartheta$  a rotation in six dimensions, but there are frequently more
than one fixed points in these cases. 
The number and localization of the fixed points depend on the details of the torus (or, equivalently,
the lattice $\Lambda$) and the space group element under consideration. There are as many inequivalent
fixed points as conjugacy classes of $S$ with non-trivial twist.

Given a space group generator $g=(\vartheta,\mu)$, it follows from eq.~\eqref{eq:gonZ} 
that the associated fixed points $z_f$ satisfy the condition
\begin{equation}
\label{eq:fixedpoints}
  g z_f = \vartheta z_f + \mu  =  z_f + \lambda_f,\qquad \lambda_f\in\Lambda\,,
\end{equation}
where the lattice translations are needed because the identity must happen in the torus.
In order to obtain all inequivalent fixed points associated with $g$, one can take different choices 
of $\lambda_f$ and then select only those that are not related by space group elements.
We note that, by using the product rule~\eqref{eq:Sproduct}, defining $g_f= (\mathbbm1,-\lambda_f) g$ 
leads to the identity $g_f z_f = z_f$. The space group element $g_f$ is typically called the
constructing element associated with the singularity $z_f$.

Let us illustrate the fixed point structure of an orbifold by using a $\mathbb T^2/\Z2\x\Z2$ orbifold 
with roto-translations.\footnote{\label{foot:example}In terms of the classification of ref.~\cite{Fischer:2012qj}, we refer here 
to a 2D subsector of the non-local geometry $(2,5)$ of a 6D \Z2\x\Z2 toroidal orbifold. By itself, the non-orientable
geometry induced by this space group cannot be used to compactify a 6D field theory as it cannot sustain chiral fermions. 
Belonging to a larger 6D orbifold solves this issue. We thank H.P. Nilles for this observation.}
We define the orbifold through the space-group roto-translation generators $g_1=(\theta,\frac{1}{2}e_1)$ and $g_2=(\omega,\frac{1}{2}e_2)$,
where $e_{1,2}$ are the orthogonal lattice generators of the torus and the \OO6 generators are given by
\begin{equation}
\label{eq:z2xz2gens}
\theta=\left( \begin{array}{cc} 1 & 0 \\ 0 & -1 \end{array}\right)\,,\qquad
\omega=\left( \begin{array}{cc} -1 & 0 \\ 0 & 1 \end{array}\right)\,,
\end{equation}
satisfying $\theta^2=\omega^2=\mathbbm1$, such that $g_1^2$ and $g_2^2$ have a trivial action on the torus, as expected
for a \Z2\x\Z2 orbifold. Omitting the translational generators $(\mathbbm1,e_i)$ and their conjugations,
the space group comprises the conjugacy classes of the elements $\{g_1,\;g_2,\;g_1g_2,\;\mathbbm1\}$.

\begin{figure}[!t!]
\begin{center}
\includegraphics{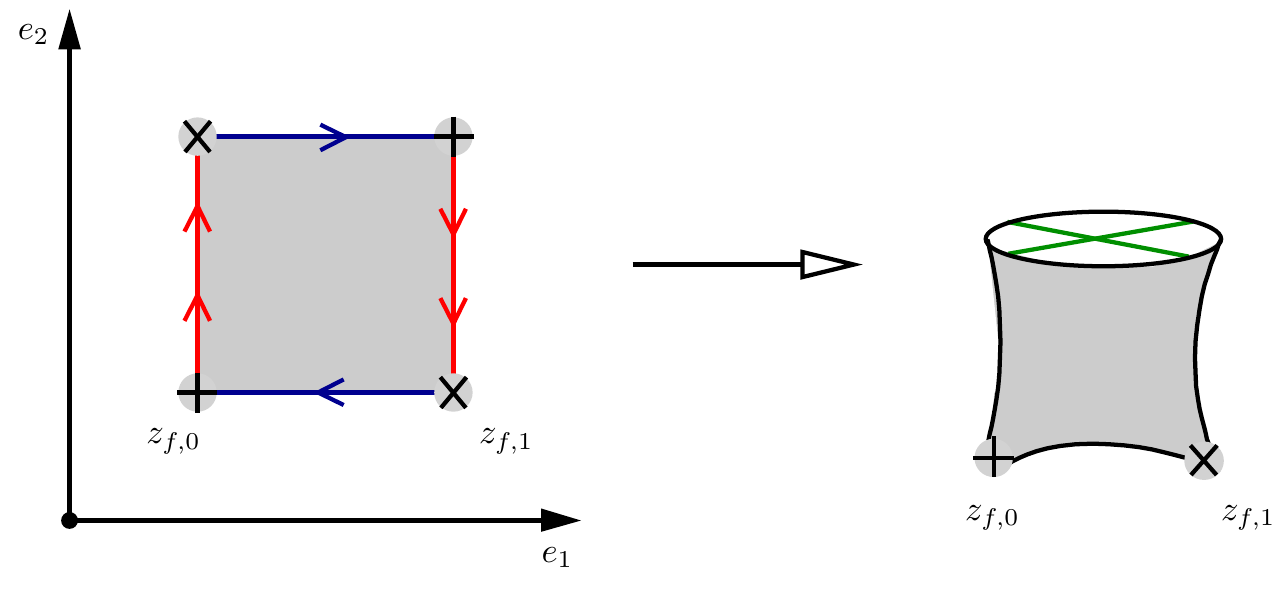}
\caption{A 2D \Z2\x\Z2 orbifold with roto-translations. The shaded region on the left corresponds to the fundamental region of the orbifold.
The fixed points $z_{f,0}$ and $z_{f,1}$ of the $g=g_1 g_2$ sector are displayed with bullets in the figure.
The fixed points at the top are identified with those in the bottom with the same symbol by the action of other
space group elements. Similar arrows are also identified. This description is equivalent to the cross-capped pillow
on the right.}
\label{fig:example}
\end{center}
\end{figure}

Let us first focus on the element $g= g_1 g_2 = (-\mathbbm1, \tfrac12(e_1-e_2))$. By applying eq.~\eqref{eq:fixedpoints},
we find four fixed points in the fundamental domain of the torus: 
$z_f\in\{\frac14(e_1+e_2),\frac14(e_1+3e_2),\frac14(3e_1+e_2),\frac34(e_1+e_2)\}$. One can easily verify that
only two of these points are inequivalent; $\frac14(e_1+e_2)$ is related to $\frac34(e_1+e_2)$ and
$\frac14(e_1+3e_2)$ is related to $\frac14(3e_1+e_2)$ in the torus by acting on them with $g_1$ or $g_2$.
Thus, one can choose the fixed points $z_{f,0}=\frac14(e_1+e_2)$ and $z_{f,1}=\frac14(3e_1+e_2)$ as the inequivalent 
fixed points associated with $g$. These points are depicted with bullets in fig.~\ref{fig:example}.
The constructing elements associated with the fixed points are given by $g_{f,0}=(-\mathbbm1,\tfrac12(e_1+e_2))$ and
$g_{f,1}=(-\mathbbm1,\tfrac12(3e_1+e_2))$.

We consider now the element $g=g_1$. For this element, it turns out that eq.~\eqref{eq:fixedpoints} has no
solution, revealing that there are no fixed points associated with this space group element. The same is true
for $g_2$. This observation will be useful when figuring out the geometric discrete symmetries of the 
compactification.

For reasons that shall be clearer in sec.~\ref{sec:heteroticorbifolds}, each set of fixed points
is named a sector. From our previous discussion, we note that, ignoring the trivial sector of the identity element,
in the $\mathbb{T}^2/\Z2\x\Z2$ orbifold worked out here there are two empty sectors and one sector with two fixed points. 
The appearance of empty sectors is related to the existence of roto-translation space group elements in toroidal Abelian 
orbifolds. In general, orbifolds without roto-translations do not exhibit empty sectors. 

The global geometric structure of the orbifold is obtained by inspecting the action of all the space group generators.
From the sector associated with $g=g_1g_2$, we find that the space group reduces the fundamental domain of $\mathbb{T}^2$ 
to $1/4$ of the torus fundamental domain, as illustrated in fig.~\ref{fig:example}. We see that the combined action of $g_1$ and $g_2$ identifies the 
singularities depicted at the top with those in the bottom, sharing the symbols $\x$ and $+$.
This crossed identification also affects the ``boundaries'' of the fundamental domain of the orbifold, which are
also identified according to the types of arrows in the figure.
From this description, we observe that this orbifold is equivalent to the well-known 2D cross-capped pillow, with p-rectangular 
Bravais lattice (see e.g. table B.2 of ref.~\cite{Fischer:2012qj}), also called singular (real) projective plane.

The structure of fixed points in an orbifold allows to determine its geometric symmetries. Notice, in our example, 
that the cross-capped pillow is symmetric under the exchange of its
inequivalent singularities. Consequently, this orbifold is invariant under $S_2\simeq\Z2$ transformations.
Analogous (permutation) $S_n$ symmetries arise in orbifolds with a different number of singularities.

\subsection{Heterotic Abelian orbifolds}
\label{sec:heteroticorbifolds}

The degrees of freedom of a string theory emerge from the left and right-moving vibrational 
modes of a string. The observation that they are independent led to the heterotic strings, which 
are the mixture of the right-moving modes, $X_R$ and $\Psi_R$, of a 10D supersymmetric string 
with the left-moving modes, $X_L$ of a 26D bosonic string. The 16 extra bosonic degrees of freedom $X_L^I$, $I=1,\ldots,16$,
are compactified on a torus $\mathbb{T}^{16}$, whose lattice vectors
are constrained by anomaly cancellation to be those of the $\Lambda_{\E8\x\E8}$ or $\Lambda_{\SO{32}}$ 
root lattice,\footnote{If one does not demand 
the resulting theory to be supersymmetric, there is a third option, the \SO{16}\x\SO{16} root lattice.}
revealing the structure of an \E8\x\E8 or \SO{32} gauge group on a 10D supersymmetric space-time.
We focus here on the heterotic string with \E8\x\E8 gauge group.

Heterotic orbifolds are constructed by compactifying six spatial dimensions of the 10D space-time
of a heterotic string on a toroidal orbifold. Right and left-moving modes, $X_R$ and $X_L$, 
mix to build the (bosonic) coordinates of the space-time, $X=X_L+X_R$, but they can still be taken
as independent degrees of freedom.
As a consequence, one can in principle choose different compactification schemes for
each mode. However, for simplicity, we focus here on so-called {\it symmetric} heterotic orbifolds,
in which both modes are compactified on the same orbifold. As already mentioned, we can also complexify 
these coordinates, so that we have two uncompactified complex dimensions, corresponding to those of 
the observed space-time, and three complex dimensions compactified on an Abelian orbifold.

Insisting on preserving $\maN=1$ supersymmetry in 4D after compactifying the $\maN=1$
heterotic strings on 6D toroidal orbifolds, restricts a number of properties of these constructions.
First, it is known that preserving $\maN=1$ requires that the point group $P$ be a subgroup of \SU3.
Recalling that the point group elements of Abelian orbifolds can be written as 
$\vartheta=\text{diag}(e^{2\pi\I v_1},e^{2\pi\I v_2}, e^{2\pi\I v_3})$, we immediately
find that the condition $v_1+v_2+v_3=0$ for each diagonalized generator leads
to obtain $\maN=1$ in 4D. Furthermore, more than two independent generators of $P$ would 
not leave any invariant supersymmetry generator; thus, only one or two distinct point group 
generators of orders $N$ and $M$ can be considered, corresponding to
cyclic \Z{N} or \Z{N}\x\Z{M} point groups. It is customary to label the 
orbifold by the name of its point group (also called {\QQ} class). In general, there
is more than one (couple of) generator(s) that can yield the same point group, but any choice
can be diagonalized in terms of the same twist vector.

Secondly, demanding that the space group elements be torus isometries further restricts both the choice of the 
tori and the space groups. For each choice of generators of a given point group, there are
different torus lattices $\Lambda$ that are left invariant under the point group. If we allow
for a number of moduli to take any values and consider the lattices so related to be equivalent, 
they build a so-called \Z{} class. Each point group admits different \Z{} classes.

Finally, once the point group and a torus lattice have been chosen for the compactification, 
one has still the freedom to consider different values of the translations $\mu$ of the space group 
generators $g=(\vartheta,\mu)$, which may be equivalent up to affine transformations or not. 
Equivalent translations together with the corresponding lattice and point group generators 
define an affine class.  

In summary, all space groups useful for orbifold compactifications are obtained by classifying 
the admissible combinations of point groups and their \Z{} and affine classes.
This has been done systematically in ref.~\cite{Fischer:2012qj}, from
which we learn that there exist 138 admissible Abelian space groups for 6D supersymmetric orbifold.
All possible point groups with their corresponding twist vectors
and the number of compatible \Z{} and affine classes are listed in table~\ref{tab:allAbelianOrbifolds}.
We shall explore the phenomenology of all 138 space groups. 

\begin{table}[!t!]
\centering
\begin{tabular}{|c|c|c|c|}
\hline
Orbifold        & Twist                                         & $\#$ of      & $\#$ of affine \\
label           & vector(s)                                     & \Z{} classes & classes\\
\hline
\hline
$\Z{3}$         & $\frac{1}{3}(1,1,-2)$                         &  $1$ &  $1$ \\
$\Z{4}$         & $\frac{1}{4}(1,1,-2)$                         &  $3$ &  $3$ \\
$\Z{6}$--I      & $\frac{1}{6}(1,1,-2)$                         &  $2$ &  $2$ \\
$\Z{6}$--II     & $\frac{1}{6}(1,2,-3)$                         &  $4$ &  $4$ \\
$\Z{7}$         & $\frac{1}{7}(1,2,-3)$                         &  $1$ &  $1$ \\
$\Z{8}$--I      & $\frac{1}{8}(1,2,-3)$                         &  $3$ &  $3$ \\
$\Z{8}$--II     & $\frac{1}{8}(1,3,-4)$                         &  $2$ &  $2$ \\
$\Z{12}$--I     & $\frac{1}{12}(1,4,-5)$                        &  $2$ &  $2$ \\
$\Z{12}$--II    & $\frac{1}{12}(1,5,-6)$                        &  $1$ &  $1$ \\
$\ZxZ{2}{2}$    & $\frac{1}{2}(0,1,-1)\;,\;\frac{1}{2}(1,0,-1)$ & $12$ & $35$ \\
$\ZxZ{2}{4}$    & $\frac{1}{2}(0,1,-1)\;,\;\frac{1}{4}(1,0,-1)$ & $10$ & $41$ \\
$\ZxZ{2}{6}$--I & $\frac{1}{2}(0,1,-1)\;,\;\frac{1}{6}(1,0,-1)$ &  $2$ &  $4$ \\
$\ZxZ{2}{6}$--II& $\frac{1}{2}(0,1,-1)\;,\;\frac{1}{6}(1,1,-2)$ &  $4$ &  $4$ \\
$\ZxZ{3}{3}$    & $\frac{1}{3}(0,1,-1)\;,\;\frac{1}{3}(1,0,-1)$ &  $5$ & $15$ \\
$\ZxZ{3}{6}$    & $\frac{1}{3}(0,1,-1)\;,\;\frac{1}{6}(1,0,-1)$ &  $2$ &  $4$ \\
$\ZxZ{4}{4}$    & $\frac{1}{4}(0,1,-1)\;,\;\frac{1}{4}(1,0,-1)$ &  $5$ & $15$ \\
$\ZxZ{6}{6}$    & $\frac{1}{6}(0,1,-1)\;,\;\frac{1}{6}(1,0,-1)$ &  $1$ &  $1$ \\
\hline
\end{tabular}
\caption{We list in the first column all 17 different Abelian point groups for 6D toroidal 
heterotic orbifolds that yield $\maN=1$ supersymmetric models in 4D. The second column displays 
the twist vectors associated with the point group generators. In the third and fourth columns, we 
show, respectively, the number of lattices (or \Z{} classes) and space-group translations (or affine classes)
that are compatible with each point group. The details of each space group are given in ref.~\cite{Fischer:2012qj}.} 
\label{tab:allAbelianOrbifolds}
\end{table}

Once one space group has been chosen to compactify the heterotic strings, the geometric features
of the orbifold in 6D are completely defined, and, due to the conformal structure of string theory, 
these properties determine some aspects of the spectrum of matter in the resulting 4D supersymmetric 
field theory. In particular, modular invariance of the heterotic string requires
that the orbifold action be embedded into the gauge degrees of freedom of the string. This means
that the space group must be translated into an equivalent group acting in the 16D space
associated with the gauge group, the so-called gauge twisting group.

The simplest such an embedding is defined (in the bosonic formulation) by two kinds of translations 
of the gauge degrees of freedom. The point group elements $\vartheta$
are embedded as shifts $V$, whereas the torus lattice vectors $e_i$ are embedded as 
so-called Wilson lines (WLs) $A_i$, $i=1,\ldots,6$. Let us explain the details by using
\Z{N}\x\Z{M} orbifolds as our working example. A generic space group element $g$ of a \Z{N}\x\Z{M}
orbifold with $P$ generators $\theta$ and $\omega$ can be embedded into the gauge degrees of freedom as
\begin{equation}
\label{eq:embedding}
  g = (\theta^n\omega^m, \mu_i e_i) \quad \hookrightarrow \quad V_g \equiv nV+mW+ \mu_i A_i\,,\qquad n,m\in\Z{}, \mu_i\in\mathbb{R}\,,
\end{equation}
where $V$ and $W$ are the 16D shift vectors of fractional entries that encode in the gauge group
the respective action of $\theta$ and $\omega$ in the 6D orbifold; $\mu_i$ are non-integers 
or integer numbers, depending on whether the space group element is a roto-translation or not; 
and the six WLs $A_i$ are also 16D fractional vectors. $V_g$ represents the gauge embedding 
of the space group $g$.

Under this gauge embedding, the action of a space group element is such that $z\mapsto gz$ in six of the ten dimensions of the space-time 
of the heterotic string, and the bosonic left-moving coordinates associated with the gauge degrees of freedom
of the heterotic strings are transformed according to
\begin{equation}
\label{gonGauge}
  X_L^I \longmapsto X_L^I + 2\pi V_g^I\,,\qquad I=1,\ldots,16\,.
\end{equation}

It is convenient to discuss the details of the states associated with the string excitations in the
dual momentum space. If we focus on the gauge momentum contribution to the states $|p\rangle_L$ with momentum $p$, 
its behavior under the action of a space group element is dominated by the left-moving contribution to the 
full vertex operator $\exp\{\I p\cdot X_L\}$ (see e.g. eq.~(2.5) of ref.~\cite{Kobayashi:2011cw}).
Under the action of $g$, this operator becomes $\exp\{2\pi\I p\cdot V_g\}\exp\{\I p\cdot X_L\}$,
which means that the momentum state acquires a phase under $g$,
\begin{equation}
\label{eq:gonLstate}
 |p\rangle_L \to e^{2\pi\I p\cdot V_g} |p\rangle_L\,,\qquad p\in\Lambda_{\E8\x\E8}\,,
\end{equation}
where $\Lambda_{\E8\x\E8}$ denotes the (self-dual and integer) root lattice of the \E8\x\E8 gauge group.

The gauge embedding is subject to some constraints. First, since the point group generators of a
\Z{N}\x\Z{M} orbifold satisfy $\theta^N=\mathbbm1=\omega^M$, the action of the shift vectors 
corresponding to $\theta^N$ and $\omega^M$ must be trivial in the gauge degrees of freedom. 
This implies that, according to eq.~\eqref{eq:gonLstate}, 
the shift vectors are constrained to satisfy $N V, M W\in\Lambda_{\E8\x\E8}$ because the lattice
is integer (i.e. the inner product of different lattice vectors is an integer). Secondly,
WLs must be consistent with the torus geometry and the orbifold action on it.
For a given point group generator, in general, $\vartheta e_i = \sum_j \gamma_{ij} e_j$ for some
integer coefficients $\gamma_{ij}$. This implies that the WLs must fulfill the relations 
$A_i = \gamma_{ij} A_j$ up to lattice translations in $\Lambda_{\E8\x\E8}$. The set of resulting
equations of this type can be reduced to conditions for the WLs; some of them
must vanish and other WLs $A_i$ have a non-trivial order $N_i$, such that 
$N_i A_i \in \Lambda_{\E8\x\E8}$ (without summation over $i$).

The final constraint on the gauge embedding comes from modular invariance, which is a string theoretical
requirement ensuring that the compactified field theory is anomaly free. In the most general case of Abelian
\Z{N}\x\Z{M} heterotic orbifolds, modular invariance requires that~\cite{Ploger:2007iq}
\begin{eqnarray}
\label{eq:ModInv}
N\,(V^2-v^2) = 0\mod2\,,           &\quad& N_i\,(V\cdot A_i) = 0\mod 2\,,\quad i=1,\ldots,6\,,\\
M\,(W^2-w^2) = 0\mod2\,,           &\quad& N_i\,(W\cdot A_i) = 0\mod 2\,, \nonumber\\
M\,(V\cdot W-v\cdot w) = 0\mod2\,, &\quad& N_i\  A_i^2 = 0\mod 2\,,\nonumber\\
                                   &\quad& \text{gcd}(N_i,N_j)\,(A_i\cdot A_j) = 0\mod 2\,,\quad i\neq j\,.\nonumber
\end{eqnarray}
Here we consider $\theta=\text{diag}(e^{2\pi\I v_1},e^{2\pi\I v_2}, e^{2\pi\I v_3})$ and
$\omega=\text{diag}(e^{2\pi\I w_1},e^{2\pi\I w_2}, e^{2\pi\I w_3})$ in terms of the two twist vectors,
$v$ and $w$.

The space group together with the corresponding gauge twisting group, fulfilling all the previous 
requirements, builds up an admissible symmetric, Abelian orbifold compactification of a
heterotic string.

The properties of the space group and a compatible gauge twisting group completely determine the matter content 
of the emerging 4D field theory. The matter fields in a heterotic orbifold correspond to the quantum states
of (left and right-moving) {\it closed} string modes, that are left invariant under the action of all
elements of the space and gauge twisting groups. String modes that are not invariant under the orbifold
do not build admissible states of the compactification. Closed strings in an orbifold are of two kinds: untwisted
and twisted strings. Untwisted strings are closed strings found among the original strings of the 10D heterotic theory
and that are not projected out by the orbifold action. Twisted strings are special. They arise only {\it because}
of the appearance of the orbifold singularities and are thus attached to them.

As in the uncompactified theory, 4D effective states consist of a left and a right-moving component. Both components must fulfill the 
so-called level-matching condition, $M_R = M_L$, whose origin is that there is no preferred point on a closed string.
For non-zero masses of string states are few times the string scale $M_{s}$, which is close to
the Planck scale, any massive state is too massive to be observed at low energies and, therefore, decouples
from the observable matter spectrum of the compactification. In string compactifications aiming at reproducing
the physics of our universe, one must thus focus on the study of massless (super)fields, $M_L=M_R=0$.

Since in 10D the only massless closed strings found in the heterotic theory are those corresponding to the \E8\x\E8 
superfields and the gravity supermultiplet, the untwisted closed-string states that are invariant under the 
orbifold represent first the unbroken 4D gauge superfields that generate the unbroken gauge group 
$\maG_{4D}\subset\E8\x\E8$, and the 4D gravity multiplet. Additionally, 
they correspond to the (untwisted) moduli, which parametrize the size and shape of the orbifold, and some 
(untwisted) matter fields that transform non-trivially under $\maG_{4D}$. The gauge properties of 
heterotic string fields are determined by their left-moving momentum, which for untwisted fields is
just a vector of the root lattice of the 10D gauge group, $p\in\Lambda_{\E8\x\E8}$. Those states
whose momenta satisfy $p\cdot V = p\cdot W = p\cdot A_i = 0\mod 1$ belong to the gravity multiplet, the 
gauge multiplets or are moduli; the rest of the states have non-trivial gauge quantum numbers and
build therefore matter fields.

The twisted states correspond to closed strings whose center of mass is at the orbifold singularities.
Their left and right-moving momenta depend on the constructing element associated with the singularity 
to which they are attached, according to our discussion in sec.~\ref{sec:fixedpoints}. The matter 
spectrum of string states of an orbifold is mostly populated by twisted fields. The gauge momentum 
of a string attached to the fixed point associated with the constructing element $g$ 
is given by $p_{sh}=p + V_g$, where $V_g$ is defined in eq.~\eqref{eq:embedding}. 
The corresponding states remain in the orbifold spectrum only if they are invariant under the action of all 
centralizer elements $h\in S$, such that $[g,h]=0$. It is thus
clear that, when some WL is chosen to vanish, $A_j=0$ for some fixed $j$ (up to lattice translations), 4D matter fields
located at the singularities with constructing elements $(\theta^n\omega^m, \mu_j e_j+\sum_{i\neq j}\mu_i e_i)$ and
$(\theta^n\omega^m, \mu'_j e_j+\sum_{i\neq j}\mu_i e_i)$ are identical concerning their quantum numbers under $\maG_{4D}$,
as long as their centralizers are equivalent. 
Following the final remarks in sec.~\ref{sec:fixedpoints}, those states would nevertheless be related under
the internal geometric (permutation) symmetry of the orbifold. However, if $A_j\neq0$, $p_{sh}$ at
various singularities differ, breaking the permutation symmetry.
These are key observations to arrive at the flavor symmetries,
as we now proceed to discuss.


\section{Flavor symmetries in Abelian heterotic orbifolds}
\label{sec:symmetries}

\subsection{Symmetries from string selection rules}
\label{sec:selection}

As long as the strings are not deformed by the space-time curvature, conformal field theory (CFT)
is a useful tool to compute, for example, the amplitude of the interactions among the fields related
to the string states~\cite{Burwick:1990tu,Erler:1992gt,Choi:2007nb,Dixon:1990pc}. 
Since orbifolds are flat everywhere but at isolated points, the description
of the string dynamics is just as in the original uncompactified theory, even after compactification
on these spaces. This is a great advantage of orbifold compactifications because we must not rely on
a supergravity approximation, which might break the connection between string theory and the 4D effective
model.

In the CFT, one determines the coupling strength of interactions among, say, $r$ effective fields 
$\Phi_\ell$, $\ell=1,\ldots,r$, by computing the $r$-point correlation functions of the vertex 
operators associated with the interacting fields,
\begin{equation}
\label{eq:correlators}
 \mathcal A = \left\langle V_{-1/2}^{(1)} V_{-1/2}^{(2)} V_{-1}^{(3)} V_0^{(4)}\cdots V_0^{(r)} \right\rangle\,,
\end{equation}
where $V_{-1/2}^{(\ell)}$ denotes a fermionic vertex operator in the $(-1/2)$-ghost picture and
$V_{0,-1}^{(\ell)}$ denote bosonic vertex operators in the $0$ or $(-1)$-ghost pictures. 
The explicit expressions are written in terms of the quantum numbers of the string states 
(cf. e.g.~\cite{Kobayashi:2011cw}), revealing that there is a number of conditions that
those quantum numbers must satisfy in order for the interaction amplitudes~\eqref{eq:correlators} 
to be non-vanishing. These conditions are known
as selection rules~\cite{Hamidi:1986vh,Dixon:1986qv,Casas:1991ac,Kobayashi:1991rp,Kobayashi:2011cw,Nilles:2013lda,Bizet:2013gf}.
The selection rules, beside gauge invariance, include $R$-charge conservation and
space-group invariance, which deserve a discussion because they lead to discrete symmetries
that may be important for flavor physics.

\paragraph{$R$-charge conservation.}
In addition to the left-moving momentum $p_{sh}$ that contains the information about
its gauge charges, a string state has the so-called $H$-momentum $q_{sh}$ in the three 
compactified, complex dimensions $z^a$. In the bosonic formulation of the heterotic string, the
entries of the $H$-momentum are fractional numbers that depend on whether they correspond 
to the description of a fermion or
a boson, differing by $\pm1/2$ units. This momentum, together with the number of left and right-moving 
oscillator perturbations acting on the vacuum, 
build the so-called $R$-charge (see e.g.~\cite{Nilles:2013lda,Bizet:2013gf}), which,
in contrast to pure $H$-momentum, is invariant under the ghost picture-changing operation.\footnote{
The ghost picture or ghost charge of the vertex operators is given as subindex in eq.~\eqref{eq:correlators}.
The total ghost charge must be $-2$ to cancel the ghost charge $+2$ of the sphere
on which $\mathcal A$ is computed. However, all different ghost-charge assignations or pictures 
yielding the same total ghost charge provide equivalent results. Thus,
it is natural to demand that physical charges be invariant under ghost-picture changing.}

By computing CFT correlation functions~\eqref{eq:correlators}, one
can demonstrate that weakly-coupled strings interact only if the total $R$-charge
of the coupling satisfies a conservation principle stated as~\cite{Nilles:2013lda,Bizet:2013gf}
\begin{equation}
\label{eq:Rconservation}
  \sum_{\ell=1}^r R_a^{(\ell)}= -1\mod N_a\,,\qquad a=1,2,3\,,
\end{equation}
where each integer $N_a$ denotes the order of the point group generators acting on the $a$-th complex 
coordinate $z^a$ of the 6D torus, i.e. such that $N_a v_a\in\Z{}$. 
If one normalizes the charges $R_a^{(\ell)}$ to be integers by multiplying by $N_a$, 
eq.~\eqref{eq:Rconservation} provides the discrete symmetry group $\Z{N_1^2}\x\Z{N_2^2}\x\Z{N_3^2}$.

On the other hand, since these $R$-charges distinguish the bosonic and fermionic components of the
4D effective superfields, the discrete symmetry arising from this invariance principle can be only
an $R$ symmetry, explaining why they are called $R$-charges. We assume here that flavor symmetries are not $R$ symmetries, 
thus the discrete, $\Z{N_1^2}\x\Z{N_2^2}\x\Z{N_3^2}$ symmetry of $R$-charges cannot
be part of a flavor symmetry.

\paragraph{Space-group invariance.}
In the compactified theory, interactions must be invariant under the space group
that defines the orbifold compactification. This implies that the joint action of 
the composition of the constructing elements of the interacting strings must be trivial
{\it on the orbifold} (rather than on the torus). This condition is the so-called
space-group selection rule. If we denote the constructing element of the fixed point 
$z_{f,\ell}$ of the sector $(\vartheta_\ell,\mu^{(\ell)})$
as $g_{f}^{(\ell)}=(\vartheta_{\ell},\, \mu_f^{(\ell)})$, the space-group selection 
rule is given by
\begin{equation}
\label{eq:spacegroupSel}
  \prod_{\ell=1}^r g_{f}^{(\ell)} = \prod_{\ell=1}^r \left(\vartheta_{\ell},\, \mu_f^{(\ell)}\right) \stackrel{!}{=} 
  \Big(\mathbbm{1},\,\bigcup_\ell \tilde\Lambda_\ell\Big)\,,\qquad 
  \tilde\Lambda_\ell=(\mathbbm{1}-\vartheta_{\ell})\Lambda\,,
\end{equation}
where e.g. $\vartheta_{\ell}=\theta^{q_{\ell}}\omega^{w_{\ell}}$ for \Z{N}\x\Z{M} orbifolds
and $\tilde\Lambda_\ell$ denotes the invariant sublattice of fixed points. The invariant sublattice of
fixed points is such that, if the fixed point $z_{f,\ell}$ has constructing element $g_{f}^{(\ell)}$
and $\tilde\lambda^{(\ell)}= (\mathbbm{1}-\vartheta_{\ell})\lambda$ with arbitrary $\lambda\in\Lambda$,
then $z_{f,\ell}+\lambda$ is the fixed point associated with the constructing element
$(\vartheta_{\ell},\, \mu_f^{(\ell)}+\tilde\lambda^{(\ell)})$ which is in the conjugacy class of $g_{f}^{(\ell)}$
and refers thus to the same fixed point in the orbifold.

In order to satisfy the space-group selection rule, we must impose first that $\prod_\ell \vartheta_\ell \stackrel{!}{=}\mathbbm1$,
which for \Z{N}\x\Z{M} orbifolds amounts to demanding
\begin{equation}
\label{eq:pointgroupSel}
  \sum_{\ell=1}^r q_\ell \stackrel{!}{=} 0\mod N\,,\qquad
  \sum_{\ell=1}^r w_\ell \stackrel{!}{=} 0\mod M\,.
\end{equation}
These relations suggest that the effective fields $\Phi_\ell$ can be considered to transform
under a discrete symmetry $\Z{N}\x\Z{M}$ with charges $(q_\ell,w_\ell)$. Nonetheless,
as we shall shortly see, these two symmetries are not always independent, yielding sometimes 
a smaller symmetry.

The second part of the space-group selection rule can be rewritten as
\begin{equation}
\label{eq:transSel}
  \mu_f^{(1)} + \sum_{\ell=2}^{r} \left( \prod_{\ell'=1}^{\ell-1} \vartheta_{\ell'} \right) \mu_f^{(\ell)} 
  \stackrel{!}{=} \sum_{\ell=1}^r \tilde\lambda^{(\ell)}\,,\qquad
  \tilde\lambda^{(\ell)} \in \tilde\Lambda_\ell\,.
\end{equation}
Since all vectors $\tilde\lambda^{(\ell)}$ and $\mu_f^{(\ell)}$ can be expressed in terms of the basis
vectors $e_i$, $i=1,\ldots,6$, eq.~\eqref{eq:transSel} becomes a set of (up to) six independent
conditions similar to those of eq.~\eqref{eq:pointgroupSel}, which depend on the specifics of
the space group elements. I.e. the 4D fields  are charged under additional \Z{N_i}, $i=1,\ldots,6$,
that depend on the space group.

To illustrate the conditions that follow from eq.~\eqref{eq:transSel}, let us consider the 
$\mathbb{T}^2/\Z2\x\Z2$ orbifold with the point-group generators given by eq.~\eqref{eq:z2xz2gens}, ignoring
the rest of the 6D heterotic orbifold (see footnote~\ref{foot:example}). A generic element
$\lambda\in\Lambda$ is written as $\lambda = \lambda_1 e_1 + \lambda_2 e_2$ with $\lambda_i\in\Z{}$.
Let us suppose that we are considering couplings among states arising only from the sector $(\theta\omega,\mu)=(-\mathbbm1,\frac12(e_1-e_2))$
since no fixed points appear in the $g_1$ and $g_2$ sectors of this orbifold.
Because of eq.~\eqref{eq:pointgroupSel} and $q_\ell = w_\ell=1$ for all massless twisted states we consider, 
we learn that the number $r$ of fields that an admissible coupling 
can have in this orbifold is even. A general element of the corresponding invariant sublattice is
given by $(\mathbbm1-\vartheta)\lambda = 2\lambda = 2\lambda_1 e_1 + 2 \lambda_2 e_2$. 
Thus, we see that eq.~\eqref{eq:transSel} takes the form
\begin{equation}
\label{eq:transSelZ2xZ2a}
  \mu^{(1)} - \mu^{(2)} + \mu^{(3)} - \mu^{(4)} + \ldots - \mu^{(r)}
  \stackrel{!}{=} 2\lambda\,,
\end{equation}
where the sign in the last vector is a consequence of $r$ being even. 
Rewriting the constructing elements as $g_{f,0}=(-\mathbbm1,\mu + e_2)$ and $g_{f,1}=(-\mathbbm1,\mu+e_1+e_2)$, so
that the field $\Phi_\ell$ in a coupling may have the constructing element
$g_{f}^{(\ell)} = (-\mathbbm1,\mu + n_1^{(\ell)} e_1 + n_2^{(\ell)}e_2)$ with $(n_1^{(\ell)},n_2^{(\ell)})=(0,1)$ or $(1,1)$,
we find that eq.~\eqref{eq:transSelZ2xZ2a} yields two (apparently) independent conditions
\begin{eqnarray}
\label{eq:transSelZ2xZ2}
  n_1^{(1)} + n_1^{(2)} + n_1^{(3)} + n_1^{(4)} + \ldots + n_1^{(r)} &\stackrel{!}{=}& 0\mod 2\,,\\
\nonumber
  n_2^{(1)} + n_2^{(2)} + n_2^{(3)} + n_2^{(4)} + \ldots + n_2^{(r)} &\stackrel{!}{=}& 0\mod 2\,,
\end{eqnarray}
where we have used that the integer $\lambda_i$ can be replaced by
$\lambda_i' - n_i^{(2)} - n_i^{(4)} - \ldots - n_i^{(r)}$, $i=1,2$, without loss
of generality. Another important observation is that the orbifold sectors corresponding to the
generators $g_1$ and $g_2$ do not lead to fixed points. This implies that there are no massless twisted
states related to those sectors. Thus, if one focuses on massless twisted states, our previous 
considerations are enough to arrive at the flavor symmetry in the effective theory.

From our discussion, one could be tempted to conclude that the discrete symmetry emerging from the 
space group is $\Z2^4$. This is wrong.
The correct discrete symmetry that massless states support is only a \Z2\x\Z2. The reason is as follows. 
First, since the point-group charges of these states satisfy $q_\ell=w_\ell=1$, if $\sum_\ell q_\ell = r =0\mod2$,
then $\sum_\ell w_\ell = r =0\mod2$ too. That is, we obtain only one independent \Z2 from these conditions.
Similarly, the second eq. of~\eqref{eq:transSelZ2xZ2} is automatically fulfilled once $r=0\mod2$ has been 
imposed because we have chosen $(n_1^{(\ell)},n_2^{(\ell)})=(0,1)$ or $(1,1)$. However, there exists 
a non-trivial condition yielding a \Z2 that does contribute to the flavor symmetry of massless states in
the sample $\mathbb{T}^2/\Z2\x\Z2$ orbifold that we study here.\footnote{It is possible
to show that $n_1^{(\ell)}+n_2^{(\ell)}$ is the only \Z2 charge that is independent of
the choice of the constructing elements we take (from their conjugacy classes). We thank Patrick K.S. 
Vaudrevange for very useful discussions and insight on this topic.} 

It must be stressed that the symmetries that we have discussed are only related to massless
states. Massive string states can wind on a torus even if it has no fixed points. In our example,
this case would correspond to constructing elements such as $g_1$ or $g_2$. When all elements
of the space group are taken into account, the corresponding symmetry becomes larger and the 
charges associated with the point-group generators and translations combine. Nevertheless,
in this paper we shall only consider massless states and leave the general discussion for
future work~\cite{Ramos-Sanchez:2018edc}.

\subsection{General structure of flavor symmetries}
\label{sec:flavorgeneral}

In orbifold compactifications (of any string theory), flavor symmetries can arise from 
the properties of the space group.
In particular, in heterotic orbifolds, they emerge as a result of combining 
the geometric properties of the extra dimensions and the symmetries emerging 
from the selection rules that we examined in the previous section.

As we have illustrated in sec.~\ref{sec:fixedpoints}, if the global structure of an orbifold
contains $n$ fixed points, the compact space exhibits an $S_n$ permutation symmetry,
which indicates that geometrically all singularities are equivalent. From the
perspective of the gauge quantum numbers, 4D effective fields $\Phi_\ell$ located at the singularities 
do not display any difference as long as the gauge embeddings 
$V_g$ associated with the singularities are equal (see eq.~\eqref{eq:embedding} and final remarks
in sec.~\ref{sec:heteroticorbifolds}).
Under these conditions, the 4D twisted fields build up non-trivial $S_n$ representations.

In the case of factorizable orbifolds, i.e. when $\mathbb T^6$ can be decomposed as 
$\mathbb T^{d_1}\times\mathbb T^{d_2}\times\cdots$, each subtorus has at least a
K\"ahler modulus that allows for differences in the effective theory of the fields 
originated in different tori. Thus, considering a number of singularities in each torus,
the full permutation symmetry of the orbifold is the product
$S_{n_1}\x S_{n_2}\x\cdots$, where each factor corresponds to the permutation symmetry
among the fixed points localized at each of the various tori.

Invariance under the full permutation group holds only if all WLs have trivial values. When 
some WLs are non-trivial, (at least some) twisted states with identical gauge quantum numbers 
located at various fixed points 
get different gauge properties and some others do not change. Hence, the 4D field theory 
is not invariant under the full permutation symmetry anymore, but only under (at most)
a (permutation) subgroup thereof. Therefore, the permutation symmetry is said to be 
explicitly broken by non-trivial WLs in heterotic orbifold compactifications. The 
permutation symmetry is completely broken when all WLs have non-trivial values.

In order to identify the permutation symmetries, it is important to notice which singularities 
prevail in the global structure of the orbifold. In simple prime \Z{N} orbifolds, the 
same singularities appear in all sectors. However, in less trivial orbifolds, different sectors 
(corresponding to inequivalent space group elements) have in general different singularities. 
It is the intersection of all sectors what determine the global structure of the orbifold. This 
means that only the singularities appearing in all sectors must be regarded to determine the
permutation symmetries. These fixed points, which include points in invariant subtori
(like those of \Z2\x\Z{M} orbifolds), exhibit
equivalent centralizers and thus the associated twisted states are equal.

Both the permutation symmetry and the Abelian space-group symmetries build a large 
set of symmetry generators, usually denoted by $(S_{n_1}\x S_{n_2}\x\cdots) \cup (\Z{N_1}\x\Z{N_2}\x\cdots)$.
The multiplicative closure of the elements of this set 
constitutes the flavor symmetry perceived by the 4D effective fields. In most cases,
the product of Abelian discrete symmetries originated from the space group, $\Z{N_1}\x\Z{N_2}\x\cdots$, 
is a normal subgroup of $(S_{n_1}\x S_{n_2}\x\cdots) \cup (\Z{N_1}\x\Z{N_2}\x\cdots)$, which implies that the flavor 
group is given by $G_F=(S_{n_1}\x S_{n_2}\x\cdots)\ltimes (\Z{N_1}\x\Z{N_2}\x\cdots)$.
Only in a few cases, the resulting symmetry requires extra generators, leading to 
a symmetry that differs from this structure. This is important when non-trivial WLs are considered.

\paragraph{Flavor symmetries in orbifolds with roto-translations.}
If the generators of the space group include roto-translations, some sectors may not exhibit fixed points. 
As a consequence, no massless states can appear in
those sectors and, therefore, the sectors can be ignored to determine the flavor symmetries of the 
massless spectrum.

As an illustration, let us study again our $\mathbb T^2/\Z2\x\Z2$ example defined by the generators around eq.~\eqref{eq:z2xz2gens}.
In that case, only the sector $g=g_1 g_2$ has two inequivalent fixed points. The sectors $g_1$ and $g_2$ do not 
exhibit fixed points and thus cannot support massless states. The global geometric structure of the orbifold is 
just that of the projective plane with two singularities, allowing, in the absence of WLs, for an $S_2\simeq\Z2$ 
permutation symmetry of the twisted states. In addition, as we discussed in section~\ref{sec:selection}, 
there is a \Z2\x\Z2 symmetry due to the space group selection rule. That is, we observe that the 4D effective theory
must be invariant under the set $S_2\cup(\Z2\x\Z2)$ of symmetries. It is possible to verify that the group \Z2\x\Z2
remains invariant under $S_2$ elements, so it is a normal subgroup, which implies that the multiplicative closure of
the set of symmetries is
$S_2\ltimes(\Z2\x\Z2)\simeq D_4$. Therefore, the corresponding flavor symmetry is $G_F=D_4$, which
coincides with the emerging flavor symmetry when only one dimension is compactified on an $S^1/\Z2$ orbifold.

\subsection{Flavor symmetries of Abelian orbifolds without Wilson lines}
\label{subsec:flavorsNoWLs}

One of the outcomes of our study is a full classification of the flavor symmetries emerging from 6D
Abelian orbifold compactifications without WLs. Interestingly, these symmetries do not depend on the specific 
string theory to be compactified. They correspond to the flavor symmetries perceived by 4D massless closed-string
states attached to the orbifold singularities. Thus, without any further elements (such as D-branes, orientifolds
and open strings), the flavor symmetries we find are common to all 4D supersymmetric models arising 
from orbifold compactifications in generic points of their moduli space.

In these orbifolds, all states associated with fixed points of a
particular sector have identical gauge quantum numbers and only their localization
in different, independent tori, $T^{d_1}, T^{d_2},\ldots$, distinguish them. By 
applying the tools explained in the previous section, we determine the flavor
symmetries of all 138 admissible Abelian orbifolds. 

Our findings are presented in table~\ref{tab:flavorsNoWLs}. Following the notation of ref.~\cite{Fischer:2012qj},
we label each Abelian orbifold, presenting its point group symmetry, and, in parentheses, the labels $(i,j)$ 
of the corresponding \Z{} and affine classes, as introduced in section~\ref{sec:heteroticorbifolds}. These space group
labels are presented in the first and third columns. In the second and fourth columns of table~\ref{tab:flavorsNoWLs}
we display the corresponding flavor symmetries at massless level.

There are three space groups which do not lead to any flavor symmetries. The reason is that no fixed points and 
thus no twisted states appear in those orbifolds. Further, there are 71 orbifolds that yield only Abelian symmetries.
The origin of this simplicity in those cases is that only one fixed point is common to all sectors
and thus only one point appears in the global structure of the orbifold, avoiding permutation
symmetries. We also observe that 45 cases include $D_4$ factors, whereas 19 space groups lead to 
$\Delta(54)$ flavor-symmetry factors, three exhibit $S_4$ and only one contains $S_7$.
In some cases the structure of the flavor symmetry follows a factorizable pattern, that is,
the resulting flavor symmetry is the direct product of two or more independent non-Abelian
symmetries; see e.g. the $(8,1)$ geometry of \Z2\x\Z2 orbifolds. However, most of the 
resulting flavor symmetries are more complicated products and quotients
of several permutation and cyclic groups.

As expected from previous studies~\cite{Kobayashi:2006wq}, $D_4$ flavor factors appear
in \Z2\x\Z{M} orbifolds whereas $\Delta(54)$ is present in $\Z3\x\Z{M}$ orbifolds. However,
we see that also other symmetries arise in those cases. Thus, only the careful
study of the space groups that we carry out here reveals the flavor symmetries of the 4D
effective theories arising from orbifold compactifications.

Note that, given a \Z{N}\x\Z{M} point group, the largest flavor symmetry arises for
$(i,j)=(1,1)$, because the space groups with $i,j>1$ correspond to non-factorizable 
6D tori and/or include roto-translations. Both features
reduce the number of fixed points in the orbifold with respect to the $(1,1)$ space group,
avoiding large permutation symmetries. Yet there are two
exceptional cases: the flavor symmetries of \Z6--I and \Z{12}--I $(1,1)$ orbifolds are
smaller than those for $i>1$. This follows from the fact that the point group 
induces only a \Z3 symmetry for the twisted states due to their localization
in the $i=1$ case.

\newpage
{\footnotesize
\begin{longtable}{|cc|c |p{0.1mm}| cc|c|} 
 \caption{\label{tab:flavorsNoWLs}{Flavor symmetries of Abelian toroidal heterotic orbifolds with point groups 
 \Z{N} and  \Z{N}\x\Z{M}. In the first and third columns, we provide the point group as well as the labels $(i,j)$
 of the torus lattice and the roto-translational element, respectively, according to the classification of~\cite{Fischer:2012qj}. 
 The second and fourth columns display the corresponding flavor symmetries. The flavor symmetry for the space group \Z2\x\Z4
 $(1,1)$ reported in~\cite{Pena:2012ki} differs from ours because they incorrectly divide an extra \Z2.}}
 \\
\cline{1-3}\cline{5-7}
\multicolumn{2}{|l|}{Orbifold} & {Flavor symmetry} & & \multicolumn{2}{l|}{Orbifold} & {Flavor symmetry}  \\
\hhline{===~===}
\endfirsthead
\cline{1-3}\cline{5-7}
\multicolumn{2}{|l|}{Orbifold}  & {Flavor symmetry} & & \multicolumn{2}{l|}{Orbifold} & {Flavor symmetry}  \\
\hhline{===~===}
\endhead
\cline{1-3}\cline{5-7}
 \endfoot
 \endlastfoot                                                     
$\Z2\x\Z2$  &  (1,1) & $D_4{}^6 / \Z2^4$                        &&             & (3,4) & $\Z2\x\Z4$  \\   
            &  (1,2) & $\Z2\x\Z2$                               &&             & (3,5) & $\Z2\x\Z4$   \\  
            &  (1,3) & $(D_4\x D_4\x D_4) / \Z2$                &&             & (3,6) & $\Z2\x\Z4$   \\
            &  (1,4) & --                                       &&             & (4,1) & $(D_4\x D_4\x \Z4 )/ \Z2$  \\
            &  (2,1) & $D_4{}^5 / \Z2^3$                        &&             & (4,2) & $\Z2\x\Z4$   \\    
            &  (2,2) & $\Z2\x\Z2$                               &&             & (4,3) & $\Z2\x\Z4$      \\
            &  (2,3) & $(D_4\x D_4\x D_4)/ \Z2^2$               &&             & (4,4) & $\Z2\x\Z4$      \\
            &  (2,4) & $\Z2\x\Z2$                               &&             & (4,5) & $\Z2\x\Z4$      \\
            &  (2,5) & $(D_4\x D_4)/\Z2$                        &&             & (5,1) & $(D_4\x D_4\x D_4\x\Z4)/ \Z2^2$   \\
            &  (2,6) & --                                       &&             & (5,2) & $\Z2\x\Z4$      \\      
            &  (3,1) & $(D_4\x D_4\x D_4\x D_4)/ \Z2^2$         &&             & (6,1) & $(D_4\x D_4\x D_4 \x\Z4 ) / \Z2^2$  \\ 
            &  (3,2) & $\Z2\x\Z2$                               &&             & (6,2) & $\Z2\x\Z4$\\
            &  (3,3) & $(D_4\x D_4) / \Z2$                      &&             & (6,3) & $\Z2\x\Z4$\\
            &  (3,4) & --                                       &&             & (6,4) & $\Z2\x\Z4$\\                            
            &  (4,1) & $(D_4\x D_4\x D_4\x D_4) / \Z2^2$        &&             & (6,5) & $\Z2\x\Z4$\\                                                        
            &  (4,2) & $\Z2\x\Z2$                               &&             & (7,1) & ($D_4\x D_4 \x \Z4)/\Z2$  \\                                                                                   
            &  (5,1) & $(D_4\x D_4\x D_4\x D_4)/ \Z2^2$         &&             & (7,2) & $\Z2\x\Z4$\\                                   
            &  (5,2) & $\Z2\x\Z2$                               &&             & (7,3) & $\Z2\x\Z4$\\                                                                            
            &  (5,3) & $\Z2\x\Z2$                               &&             & (8,1) & $(D_4\x D_4 \x\Z4) / \Z2^2$ \\   
            &  (5,4) & $(D_4\x D_4)/\Z4$                        &&             & (8,2) & $\Z2\x\Z4$\\                                
            &  (5,5) & $\Z2\x\Z2$                               &&             & (8,3) & $\Z2\x\Z4$\\  
            &  (6,1) & $(D_4\x D_4\x D_4\x D_4) / \Z2^2$        &&             & (9,1) & $(D_4\x D_4 \x \Z4 )/ \Z2$  \\  
            &  (6,2) & $\Z2\x\Z2$                               &&             & (9,2) & $\Z2\x\Z4$\\            
            &  (6,3) & $D_4$                                    &&             & (9,3) & $\Z2\x\Z4$\\   
            &  (7,1) & $(D_4\x D_4 \x D_4)/\Z2$                 &&             & (10,1)& $\Z2\x\Z4$\\                             
            &  (7,2) & $\Z2\x\Z2$                               &&             & (10,2)& $\Z2\x\Z4$\\ 
\cline{5-7}                                                     
            &  (8,1) & $D_4\x D_4$                              && \Z2\x\Z6-I  & (1,1) & $(D_4\x D_4 \x \Z6)/\Z2$ \\                       
            &  (9,1) & $(D_4\x D_4\x D_4)/ \Z2$                 &&             & (1,2) &  $\Z2\x\Z6$\\   
            &  (9,2) & $\Z2\x\Z2$                               &&             & (2,1) & $(D_4\x D_4 \x \Z6)/\Z2$ \\  
            &  (9,3) & $\Z2\x\Z2$                               &&             & (2,2) &  $\Z2\x\Z6$\\ 
\cline{5-7}                                                     
            & (10,1) & $D_4 \x D_4$                             && \Z2\x\Z6-II & (1,1) & $\Z2\x\Z6$\\
            & (10,2) & $\Z2\x\Z2$                               &&             & (2,1) & $\Z2\x\Z6$\\
            & (11,1) & $(D_4\x D_4\x D_4)/ \Z2$                 &&             & (3,1) & $\Z2\x\Z6$\\  
            & (12,1) & $D_4\x D_4$                              &&             & (4,1) & $\Z2\x\Z6$\\
\cline{5-7}                                                     
            & (12,2) & $\Z2 \x\Z2$                              && \Z3\x\Z3    & (1,1) & $(\Delta(54)\x \Delta(54)\x \Delta(54))/ \Z3$  \\  
\cline{1-3}  
\Z2\x \Z4   &  (1,1) & $(D_4\x D_4\x D_4\x D_4 \x\Z4) / \Z2^3$  &&             & (1,2) &  $\Z3\x\Z3$         \\
            &  (1,2) & $\Z2\x\Z4$                               &&             & (1,3) &  $\Z3\x\Z3$ \\
            &  (1,3) & $\Z2\x\Z4$                               &&             & (1,4) &  $(\Delta(54) \x \Delta(54))/\Z3$  \\
            &  (1,4) & $\Z2\x\Z4$                               &&             & (2,1) &  $\Delta(54) \x \Delta(54))$  \\
            &  (1,5) & $\Z2\x\Z4$                               &&             & (2,2) &  $\Z3\x\Z3$ \\
            &  (1,6) & $(D_4\x D_4\x D_4 \x\Z4 ) / \Z2^2$       &&             & (2,3) &  $\Z3\x\Z3$      \\  
            &  (2,1) & $(D_4\x D_4\x D_4\x D_4 \x\Z4) / \Z2^3$  &&             & (2,4) &  $(\Delta(54) \x \Delta(54))/\Z3$      \\        
            &  (2,2) & $\Z2\x\Z4$                               &&             & (3,1) &  $ \Delta(54) \x \Delta(54)$  \\   
            &  (2,3) & $\Z2\x\Z4$                               &&             & (3,2) &  $\Z3\x\Z3$ \\
            &  (2,4) & $(D_4\x D_4\x \Z4)/\Z2$                  &&             & (3,3) &  $(\Delta(54) \x \Delta(54))/\Z3$ \\                             
            &  (2,5) & $\Z2\x\Z4$                               &&             & (4,1) &  $\Delta(54) \x \Delta(54)$ \\                                                                  
            &  (2,6) & $\Z2\x\Z4$                               &&             & (4,2) &  $\Z3\x\Z3$       \\                                                                             
            &  (3,1) & $(D_4\x D_4 \x D_4 \x\Z4) / \Z2^2$       &&             & (4,3) &  $(\Delta(54) \x \Delta(54))/\Z3$    \\                                       
            &  (3,2) & $\Z2\x\Z4$                               &&             & (5,1) &  $\Z3\x\Z3$ \\
            &  (3,3) & $\Z2\x\Z4$                               &&             &       &       \\   
$\Z3\x\Z6$  &  (1,1) & $\Delta(54)\x \Z6$                       && $\Z3$       & (1,1) & $(\Delta(54)\x \Delta(54) \x \Delta(54) )/\Z3^2$\\   
\cline{5-7}
            &  (1,2) & $\Z6\x\Z3$                               && $\Z4$       & (1,1) & $(D_4\x D_4 \x D_4 \x D_4 \x \Z4) / \Z2^4$ \\      
            &  (2,1) & $\Delta(54)\x \Z6$                       &&             & (2,1) & $(S_4\x S_2 \x S_2) \ltimes (\Z4^3\x\Z2^3)$  \\
            &  (2,2) & $\Z6\x\Z3$                               &&             & (3,1) & $(S_4\x S_4)\ltimes(\Z4^5\x\Z2^2)$  \\   
\cline{1-3} \cline{5-7}             
$\Z4\x\Z4$  &  (1,1) & $(D_4\x D_4\x D_4 \x \Z4\x\Z4 )/ \Z2^3$  && $\Z6$-I     & (1,1) & $\Delta(54)$ \\ 
            &  (1,2) & $\Z4\x\Z4$                               &&             & (2,1) & $(\Delta(54)\x \Z6)/\Z3$  \\
\cline{5-7}
            &  (1,3) & $\Z4\x\Z4$                               && $\Z6$-II    & (1,1) & $\Delta(54)\x [D_4 \x D_4/\Z2]$\\
            &  (1,4) & $\Z4\x\Z4$                               &&             & (2,1) & $[(\Delta(54)\x \Z6)/\Z3]\x[D_4{}^2/\Z2^2]$ \\
            &  (2,1) & $(D_4\x D_4 \x \Z4^2) / \Z2^2$           &&             & (3,1) & $[(\Delta(54)\x \Z6)/\Z3]\x[D_4{}^2/\Z2^2]$\\     
            &  (2,2) & $\Z4\x\Z4$                               &&             & (4,1) & $[(\Delta(54)\x \Z6)/\Z3] \x [D_4/\Z2]$\\
\cline{5-7}
            &  (2,3) & $\Z4\x\Z4$                               && $\Z7$       & (1,1) & $S_7 \ltimes \Z7^6$ \\
\cline{5-7}
            &  (2,4) & $\Z4\x\Z4$                               && $\Z8$-I     & (1,1) & $(D_4 \x D_4 \x \Z8)/ \Z2^2 $  \\  
            &  (3,1) & $(D_4 \x D_4\x\Z4^2) / \Z2^2$            &&             & (2,1) & $(D_4 \x D_4 \x \Z8)/ \Z2^2 $  \\ 
            &  (3,2) & $\Z4\x\Z4$                               &&             & (3,1) & $S_4 \ltimes (\Z8\x\Z4^2\x \Z2)$ \\ 
\cline{5-7}
            &  (4,1) & $(D_4\x D_4 \x \Z4^2) / \Z2^2$           && $\Z8$-II    & (1,1) & $(D_4 \x D_4 \x D_4 \x \Z8) / \Z2^3$ \\
            &  (4,2) & $\Z4\x\Z4$                               &&             & (2,1) & $(D_4 \x D_4 \x \Z8)/ \Z2^2 $ \\
\cline{5-7}
            &  (4,3) & $\Z4\x\Z4$                               && \Z{12}-I    & (1,1) & $\Delta(54)$   \\   
            &  (5,1) & $\Z4\x\Z4$                               &&             & (2,1) & $(\Delta(54) \x \Z{12}) / \Z3 $ \\   
\cline{5-7}
            &  (5,2) & $\Z4\x\Z4$                               && \Z{12}-II   & (1,1) & $(D_4\x D_4)/ \Z2$  \\ 
\cline{1-3}     
$\Z6\x\Z6$  &  (1,1) & $\Z6\x\Z6$                               &&             &       & \\
\cline{1-3}\cline{5-7}
\end{longtable} 
}

One of the conclusions one may draw from these results is that the 4D massless spectrum of 
supersymmetric heterotic orbifold compactifications can only have one of the flavor 
symmetries presented here or a subgroup thereof. However, these symmetries may
be enhanced by imposing very special conditions on the vacuum, that is, by 
requiring that the expectation values of moduli satisfy particular relations.
E.g., if one demanded that all K\"ahler moduli of the $(1,1)$ case of \Z3\x\Z3 have 
the same value, the $\Delta(54)^3/\Z3$ flavor symmetry would be enhanced to the 
multiplicative closure of $S_{27}\cup\Z3{}^5$.

It has also been shown that the discrete flavor symmetries found here can be enlarged to continuous 
gauge symmetries at some symmetry-enhanced points of the moduli space~\cite{Beye:2014nxa}. In this work, 
as already pointed out, we suppose that, if moduli stabilization is possible in these scenarios~\cite{Parameswaran:2010ec,Dundee:2010sb},
the vacua obtained correspond in general to non-enhanced points in the moduli space.

\section{\Z{N}\x\Z{M} heterotic orbifolds with MSSM-like properties}
\label{sec:class}

The flavor symmetries classified and showed in table~\ref{tab:flavorsNoWLs} correspond to the largest
symmetries that the 4D effective field models emerging from orbifold compactifications exhibit.
However, in general, they are not the flavor symmetries that models with semi-realistic features 
have because those models include non-trivial WLs, which break the permutation symmetries and
thus the flavor symmetries.

To determine the unbroken flavor symmetries that promising models can have, one must know the WL-structure
of {\it all} promising orbifold compactifications in the different geometries, and, furthermore, 
the unbroken subgroup of the flavor groups once non-trivial WLs are included in those orbifold
geometries. 

Clearly, performing a full classification of semi-realistic heterotic orbifolds is beyond 
our capabilities. The first reason is that, given the vastness of the landscape, 
even the best available algorithms to look for promising models could miss some of them. 
A second reason is that any such a classification will certainly be very time-consuming.
Instead, we use the \texttt{orbifolder}~\cite{Nilles:2011aj} to perform a random search of 
Abelian orbifold compactifications with properties similar to those of the MSSM. Even in this context,
exploring all geometries is very challenging. Thus, since it seems more likely to find
appropriate phenomenology with non-Abelian flavor symmetries, we restrict ourselves to a search of
phenomenologically promising models, considering only the 64 (19 \Z{N} and 45 \Z{N}\x\Z{M})
orbifold geometries that allow for non-Abelian flavor symmetries in the absence of WLs 
(see table~\ref{tab:flavorsNoWLs}).

\begin{table}[!b!]
\centering
\begin{tabular}{|cr|c|R{8mm}R{8mm}R{8mm}R{8mm}|r|}
\hline
 \multicolumn{2}{|c|}{}        & \multicolumn{1}{c|}{Max \#  of } & \multicolumn{4}{c|}{\#  of MSSM-like models with} &\multirow{3}{*}{Total}\\
 \multicolumn{2}{|c|}{Orbifold}& \multicolumn{1}{c|}{independent} & 0    & 1   & 2   & 3                              &                      \\
 \multicolumn{2}{|c|}{}        & \multicolumn{1}{c|}{WLs}         & \multicolumn{4}{c|}{ vanishing WL}                &                      \\
\hline
\hline
\Z4       & (2,1) & 3 &   149 &    0 &  0 & 0 &   149\\
          & (3,1) & 2 &    27 &    0 &  0 &   &    27\\
\hline
\Z6-I     & (1,1) & 1 &    30 &    0 &    &   &    30\\
          & (2,1) & 1 &    30 &    0 &    &   &    30\\
\hline
\Z6-II    & (1,1) & 3 &    26 &  337 &  0 & 0 &   363\\
          & (2,1) & 3 &    14 &  335 &  0 & 0 &   349\\
          & (3,1) & 3 &    18 &  335 &  0 & 0 &   353\\
          & (4,1) & 2 &    44 &  312 &  0 &   &   356\\ 
\hline
\Z7       & (1,1) & 1 &     1 &    0 &    &   &     1\\
\hline
\Z8-I     & (1,1) & 2 &   230 &  38 &   0 &   &   268\\
          & (2,1) & 2 &   205 &  41 &   0 &   &   246\\
          & (3,1) & 1 &   389 &   0 &     &   &   389\\ 
\hline
\Z8-II    & (1,1) & 3 & 1,604 & 398 &  21 & 0 & 2,023\\
          & (2,1) & 2 &   274 & 231 &   0 &   &   505\\  
\hline
\Z{12}-I  & (1,1) & 1 &   556 &   0 &     &   &   556\\
          & (2,1) & 1 &   555 &   0 &     &   &   555\\ 
\hline
\Z{12}-II & (1,1) & 2 &   279 &  84 &   0 &   &   363\\
\hline 
\multicolumn{8}{c}{}\\
\end{tabular}
\caption{\label{tab:ZNclassification} Number of \Z{N} heterotic orbifold models with different geometries yielding the MSSM matter spectrum. 
In the first column it is shown the orbifold label according to \cite{Fischer:2012qj}.
The maximum number of independent WLs is written in the second column and the number of models found for each number of vanishing
WLs is also shown. In the final column we display the total number of MSSM-like models.}
\end{table} 

We shall regard here an orbifold compactification as phenomenologically viable 
if its 4D effective massless spectrum satisfies the following requirements:
\begin{itemize}
\item the unbroken gauge group is $\maG_{SM}\x\maG_\text{hidden} = \SU3_c\x\SU2_L\x\U1_Y \x \maG_\text{hidden}$,
where $\maG_\text{hidden}$ contains additionally (Abelian and non-Abelian) continuous gauge factors, and the 
$\U1_Y$ is non-anomalous and compatible with grand unification;
\item the effective (twisted and untwisted) states include fields that reproduce the matter spectrum of
the MSSM; and
\item additional effective states are vector-like with respect to $\maG_{SM}$ and include 
SM singlets that can play the role of right-handed neutrinos.
\end{itemize}

With these restrictions, we have performed a broad (although non-exhaustive) search of inequivalent 
promising models arising from Abelian toroidal orbifold geometries that exhibit non-Abelian flavor symmetries 
in the absence of WLs. We have studied 19 geometries of \Z{N} orbifolds and 45 geometries 
of \Z{N}\x\Z{M} orbifolds, including cases with roto-translations.
In tables~\ref{tab:ZNclassification} and~\ref{tab:ZNxZMclassification}, we report the results of our search. 

\begin{table}[!b!]
\centering
{\small
\begin{tabular}[T!]{|cr|c|R{9mm}R{9mm}R{9mm}R{9mm}R{9mm}|r|}
\hline
\multicolumn{2}{|c|}{}         & Max \#  of    & \multicolumn{5}{c|}{\#  of MSSM-like models with} & \multirow{3}{*}{Total}   \\
\multicolumn{2}{|c|}{Orbifold} & independent   & 0    & 1   &  2  & 3  &  $\geq4$                  &         \\
\multicolumn{2}{|c|}{}         & WL            &  \multicolumn{5}{c|}{ vanishing WL}               &  \\
\hline
\hline
$\Z{2}\x \Z{2}$ & (1,1)& 6 &      1 &   152 &    52 &  0 & 0 &    205\\
                & (2,1)& 5 &     13 &   342 &    14 &  0 & 0 &    369\\
                & (3,1)& 5 &      4 &   400 &    40 &  0 & 0 &    444\\
                & (5,1)& 4 &      2 &    40 &     0 &  0 & 0 &     42\\
                & (6,1)& 4 &    344 &    57 &     0 &  0 & 0 &    401\\
                & (7,1)& 4 &     21 &    55 &     0 &  0 & 0 &     76\\
                & (8,1)& 4 &     25 &     0 &     0 &  0 &   &     25\\
                & (9,1)& 3 &     25 &     2 &     0 &  0 &   &     27\\
                &(10,1)& 3 &     19 &     2 &     0 &  0 &   &     21\\
                &(12,1)& 2 &      3 &     0 &     0 &    &   &      3\\
\hline
$\Z{2}\x \Z{4}$& (1,1) & 4 &    454 & 8,637 & 1,463 & 26 & 0 & 10,580\\  
               & (1,6) & 2 &     65 &    21 &     0 &    &   &     86\\
               & (2,1) & 4 &    260 & 4,686 & 1,131 & 81 & 0 &  6,158\\
               & (2,4) & 2 &    281 &    47 &     0 &    &   &    328\\
               & (3,1) & 3 & 18,440 & 3,762 &   103 &  0 &   & 22,305\\  
               & (4,1) & 3 &  2,911 & 1,575 &    33 &  0 &   &  4,519\\  
               & (5,1) & 3 &  1,311 &   742 &    63 &  0 &   &  2,116\\
               & (6,1) & 3 &  1,814 & 1,374 &    58 &  0 &   &  3,246\\
               & (7,1) & 3 &  1,481 & 1,122 &    64 &  0 &   &  2,667\\
               & (8,1) & 2 &    839 &    72 &     0 &    &   &    911\\
               & (9,1) & 2 &  1,620 &   522 &     0 &    &   &  2,142\\
\hline
\Z2\x\Z6-I     & (1,1) & 2 &    467 &   116 &     0 &    &   &    583\\ 
               & (2,1) & 2 &    275 &    78 &     0 &    &   &    353\\ 
\hline
\Z3\x\Z3       & (1,1) & 3 &    40 &    987 &    81 &  0 &   &  1,108\\ 
               & (1,4) & 1 &     8 &      0 &       &    &   &      8\\ 
               & (2,1) & 2 & 1,713 &    239 &     0 &    &   &  1,952\\
               & (3,1) & 2 &     6 &      0 &     0 &    &   &      6\\
               & (4,1) & 2 &   105 &    110 &     0 &    &   &    215\\ 
\hline
\Z3\x\Z6       & (1,1) & 1 & 4,469 &     24 &       &    &   &  4,493\\ 
               & (2,1) & 1 &   495 &     45 &       &    &   &    540\\
\hline
\Z4\x\Z4       & (1,1) & 3 & 1,509 & 24,693 & 2,442 &  5 &   & 28,649\\  
               & (2,1) & 2 & 6,286 &  3,548 &    19 &    &   &  9,853\\  
               & (3,1) & 2 & 4,513 &  1,003 &     6 &    &   &  5,522\\  
               & (4,1) & 2 & 3,097 &  1,627 &     6 &    &   &  4,730\\  
\hline
\hline
\Z6\x\Z6       & (1,1) & 0 & 3,696 &        &       &    &   &  3,696\\  
\hline
\multicolumn{9}{c}{} \\
\end{tabular}}
\caption{\label{tab:ZNxZMclassification} Number of \Z{N}\x\Z{M} heterotic orbifold models with different geometries yielding the MSSM matter spectrum. 
The maximum number of independent WLs and the total number of models is also shown. The number of promising models of \Z6\x\Z6 orbifolds is presented 
as a representative of geometries not admitting WLs and yielding 4D field theories endowed only with Abelian flavor symmetries; 
it is remarkable to find a large number of semi-realistic models even without WLs.}
\end{table}

Table~\ref{tab:ZNclassification} displays the number of \Z{N} orbifolds models with promising features. There are 
no models with the required properties of \Z3 and \Z4 $(1,1)$ orbifolds, reason why these cases are
not presented. In the first column, we label each orbifold geometry as in table~\ref{tab:flavorsNoWLs}.
In the second column, we write the maximal number of admissible WLs. The numbers presented in 
the third through sixth columns correspond to the number of models with 0, 1, 2 and 3 {\it vanishing} 
WLs. For example, we have found 398 semi-realistic \Z8-II $(1,1)$ orbifold models with one vanishing WL       
(out of maximally three possible WLs), i.e. there are 398 promising models with two                           
{\it non-zero} WLs. The last column provides the total number of models of each geometry.
In total, we find 6,563 phenomenologically viable models arising from all \Z{N} orbifold geometries.          
We notice that about 52\% of these models arise from the different geometries of \Z8 orbifolds.               

In table~\ref{tab:ZNxZMclassification}, where the same notation as in table~\ref{tab:ZNclassification} is followed,
we show our results for the \Z{N}\x\Z{M} 
geometries we selected because they exhibit non-Abelian flavor symmetries in the absence of
WLs. Only space groups that yield promising models are presented; this is why only
34 (out of the 45 chosen) geometries are listed.

Some \Z2\x\Z2 orbifold geometries cannot produce MSSM-like models because their
structures forbid chiral matter. This was already pointed out in~\cite{Nibbelink:2012de,Fischer:2013qza} 
as a consequence of the conditions imposed by the space group: in some \Z2\x\Z2 orbifolds, all sectors yield 
independent effective 6D $\maN=2$ supersymmetric effective theories, whose combination corresponds
to non-chiral field theories from the 4D perspective of the full compactification.

Table~\ref{tab:ZNxZMclassification} includes the results
for \Z6\x\Z6 orbifolds, even though this geometry does not provide non-Abelian 
flavor symmetries at massless level, solely for the purpose of comparison.

Excluding \Z6\x\Z6, we have found 114,683 \Z{N}\x\Z{M} heterotic orbifold compactifications              
whose 4D effective theories satisfy our phenomenological constraints. 
Interestingly, about 48\% of these promising models arise from the different geometries                  
of \Z2\x\Z4 orbifolds. 
As expected, most models with the features of the MSSM require non-trivial WLs; however, there 
are a few \Z3\x\Z6 and \Z4\x\Z4 examples where the shift vector suffices to render a
consistent gauge embedding of the compactification geometry with 4D promising features.

In comparison, we note that \Z6\x\Z6 is much more fruitful in this sense. Although \Z6\x\Z6 orbifolds
do not admit non-trivial WLs, there are thousands of models with MSSM-like properties.
This observation may trigger a phenomenological study on heterotic orbifold models with 
Abelian flavor symmetries.

On the other hand, we find that there are only 422 promising models with roto-translations,              
which can be identified from table~\ref{tab:ZNxZMclassification}, by inspecting the labels
$(i,j)$: those space groups with $j>1$ include roto-translations. One of the reasons for
this behavior is that space groups with roto-translations impose more restrictions on
the admissibility of WLs, thus making more difficult the appearance of
MSSM-like compactifications.

The list of all 121,246 \Z{N} and \Z{N}\x\Z{M} promising orbifold compactifications of the \E8\x\E8 heterotic 
string found in this study is provided in~\cite{WebTables:2018xx}, where not only the defining data
(as required by the \texttt{orbifolder}) for each of the models is provided, but also their associated flavor symmetries.
Although our results are compatible with previous findings~\cite{Nilles:2014owa,Ramos-Sanchez:2017lmj},
we find as many as nine times more models than preceding studies. Thus, our results represent the most       
exhaustive search of semi-realistic string compactifications so far.

\section{Flavor symmetries in promising string compactifications}
\label{sec:stringyFlavors}

One purpose of this work is to provide the flavor symmetries that phenomenologically viable
Abelian orbifolds admit. Since most of the promising models discussed in the preceding section
require non-zero WLs, we investigate now the flavor symmetries that
arise when non-trivial WLs are included in orbifold models with the geometries that led to 
our promising models.

As stated in section~\ref{sec:flavorgeneral}, non-zero WLs fully break some of the permutation $S_n$ 
symmetries. If there are independent permutation symmetries, different 
WLs can break them if they acquire non-trivial values. Consequently, if some WLs have non-vanishing 
values, the flavor group of the effective model is a (non-Abelian or Abelian) subgroup of the 
classified flavor symmetries of table~\ref{tab:flavorsNoWLs}.

A 6D orbifold compactification can have up to six non-trivial WLs $A_i$ of different 
orders $N_i$, but the constraints on the gauge embedding imposed by each space group,
discussed in section~\ref{sec:heteroticorbifolds}, inhibit non-trivial values
for some (and, in some cases, all) of them. For example, in the 
2D orbifold introduced in section~\ref{sec:fixedpoints}, one can verify that the WLs 
$A_1$ and $A_2$ associated with the directions $e_1$ and $e_2$ must be trivial. Thus,
if a promising model appeared from such a geometry, its flavor symmetry would then
be $D_4$. However, many geometries do admit non-trivial WLs. Details of the general
properties of the WLs allowed by all Abelian space groups are given in ref.~\cite{Fischer:2012qj}.

We have systematically determined the flavor symmetries that appear once non-trivial
WLs are included in the orbifold geometries that allow for non-Abelian flavor symmetries, 
according to table~\ref{tab:flavorsNoWLs}. Our results are presented in tables~\ref{tab:ZNflavorviable} 
for \Z{N} and~\ref{tab:ZNxZMflavorviable} for \Z{N}\x\Z{M} orbifold geometries, 
where only those space groups that allow for at least one WL are shown.
In those tables, we label the orbifold geometries according to their space groups, 
using, as before, the notation of ref.~\cite{Fischer:2012qj}. After the label, the maximal 
possible number of inequivalent non-trivial WLs is presented. 

In the fourth through seventh columns of table~\ref{tab:ZNflavorviable}, we provide the 
flavor symmetries that arise in \Z{N} orbifolds when $\ell=1,\ldots,4$ non-trivial WLs
are allowed. Since some orbifold geometries admit WLs of different orders and/or, even if
they have the same order, their action is not symmetric in all compact directions,
there may be more than one possible flavor symmetry for the same number of non-vanishing WLs. 

For example, consider the space group \Z4 $(2,1)$, that admits up to three non-trivial WLs,
two of which must have order two and one must be of order four, and yields the flavor
group $(S_4\x S_2{}^2)\ltimes(\Z4{}^3\x\Z2{}^3)$. Non-trivial values for an order-2 WL
break an $S_2$ whereas the order-4 WL breaks the $S_4$ permutation symmetry; that is, if 
one order-2 WL and one order-4 WLs are given non-trivial values, the flavor symmetry
contains only $S_2$ as permutation factor, while if both order-2 WLs acquire non-trivial values, 
only $S_4$ appears. The resulting flavor symmetries in these cases are $S_2\ltimes (\Z4\x\Z2)^2$
and $S_4\ltimes(\Z4\x\Z2)^2$, respectively. The breakdown of a \Z4\x\Z2 factor in the former 
case is related to the multiplicative closure: it is automatically broken when $S_4$ is 
no longer a symmetry. Both possible flavor symmetries with $\ell=2$ WLs are stacked one over the other in the
corresponding column of table~\ref{tab:ZNflavorviable}. We repeat this reasoning for all geometries.

Just below each flavor symmetry, we show the number of heterotic orbifold models with
phenomenologically appealing properties found in our search with such flavor symmetry (see section~\ref{sec:class})
and a given number of non-trivial WLs. There are several flavor symmetries with which 
no promising model can be associated. The final column of the table
counts the total of MSSM-like models corresponding to each space group.

Table~\ref{tab:ZNxZMflavorviable} follows a similar notation, but there are some differences. 
First, in some \Z{N}\x\Z{M} orbifold geometries, there are WLs that do not alter the degeneracy of the fixed points
even though they do have an impact on the gauge group and other 4D properties of the model.
For this reason, we provide in the fourth column the maximal number of WLs that affect the 
flavor group. Secondly, in some other cases, we find different symmetries for orbifolds with up to $\ell=6$ 
non-trivial WLs, which are given in the fifth through tenth columns. 

As before, we also provide under each flavor symmetry the number of inequivalent promising models found
with such symmetries. The total number of phenomenologically viable models is given in the last 
column. We recall here that, as we already observed in table~\ref{tab:ZNxZMclassification}, 
there are some MSSM-like models that do not require non-trivial WLs; their flavor symmetries do not 
appear in table~\ref{tab:ZNxZMflavorviable} (because they are included in table~\ref{tab:flavorsNoWLs}), 
but they are counted as part of the total number of models.

\subsection{Distribution of flavor symmetries}

Inspecting our results given in tables~\ref{tab:ZNflavorviable} and~\ref{tab:ZNxZMflavorviable} reveals 
that (excluding the 3,696 models arising from \Z6\x\Z6 orbifolds) the 121,246 promising models       
identified in the previous section have one of three types of flavor symmetries:
\begin{itemize}
\item Products and quotients of powers of $D_4$ with Abelian \Z{n} factors. We identified as many
as 91,449 models with this kind of non-Abelian flavor symmetries, which amount to about 75.4\% of    
all promising models. The most frequent combination is $D_4 \x\Z4^2\x\Z2$ , which arises naturally
in \Z4\x\Z4 orbifolds.

\item Pure Abelian flavor symmetries, including (direct) products of \Z2, \Z3, \Z4, \Z6, \Z7, \Z8 and \Z{12}
at different powers. These groups result from the breakdown of all permutation symmetries by the WLs and
are thus the symmetries arising from the space-group selection rule. 
We found 28,331 models of this type, corresponding to about 23.4\% of the total.                     

\item Products and quotients of powers of $\Delta(54)$ with Abelian \Z{n} factors. We found
only 1,466 models with these flavor symmetries, which represent about 1.2\% of all MSSM-like Abelian 
heterotic orbifolds we obtained. Models with these flavor symmetries arise only from orbifolds  
whose space group has a \Z3 generator, which are not many.
\end{itemize}
The defining parameters for our promising models, together with their flavor symmetries, are 
given in~\cite{WebTables:2018xx}.

The fact that $D_4$ appears in the majority of our models was expected because we saw already 
from table~\ref{tab:flavorsNoWLs} that most of the space groups yielding non-Abelian flavor 
symmetries in compactifications without WLs contain $D_4$. However, the proportion with respect to
models endowed with $\Delta(54)$ is much larger than expected, disfavoring somewhat the latter.

It is known that orbifold compactifications with a $D_4$ flavor symmetry and an MSSM-like matter spectrum are 
such that matter generations split in $\bs2+\bs1$ representations of $D_4$, where the third generation and its
mixings are distinct from the other two, producing some reasonable
CKM patterns once the flavor symmetry is broken by VEVs of some SM singlet fields, which turns out to be required
by moduli stabilization and decoupling of exotics. Thus, we conclude that most heterotic
orbifold compactifications with non-Abelian flavor symmetries follow these patterns, which may
deserve further study.

Even though $\Delta(54)$ is not a favored non-Abelian flavor symmetry in our constructions, the number of 
promising models is still significant and must, therefore, be considered. Phenomenologically, it has been observed that
in \Z3\x\Z3 heterotic orbifolds furnished with this flavor symmetry, SM generations frequently appear in 
flavor triplets~\cite{Carballo-Perez:2016ooy}. This means that these appealing models are endowed with three identical
SM generations, justifying the origin of the flavor multiplicity, but, even after spontaneous flavor-symmetry breaking,
complicating the explanation of the observed quark and neutrino mixing patterns. It might be interesting to
investigate whether and how this situation is improved in some other models with this symmetry. 

We finally observe that we find that almost no space group with roto-translations leads to MSSM-like models. Only the orbifold
geometries \Z2\x\Z4 $(1,6)$, \Z2\x\Z4 $(2,4)$ and \Z3\x\Z3 $(1,4)$ include roto-translations and
yield promising models, although the models of \Z3\x\Z3 $(1,4)$ admit only Abelian flavor symmetries.
One could therefore argue that roto-translations impose generally too tight constraints 
to arrive at promising Abelian orbifolds. It would be nevertheless interesting to know what kind of phenomenology
is produced by these models.

\section{Final remarks}
\label{sec:conclusions}

With the goal in mind of finding in string theory some guidance principle that singles out
the discrete symmetry that might govern the mixing patterns of fermions in the SM, we have
investigated the flavor symmetries that arise from compactifying symmetrically 
the heterotic string on Abelian toroidal orbifolds.

First, we have classified the flavor symmetries associated with the geometry of all admissible
Abelian toroidal orbifolds. This classification, presented in table~\ref{tab:flavorsNoWLs}, 
is valid for the massless closed-string sector of all string theories compactified on orbifolds
for arbitrary values of their moduli.
We find that 64 out of 138 admissible space groups yield non-Abelian flavor symmetries,
where products of $D_4$, $\Delta(54)$, $S_4$, $S_7$ and cyclic groups appear. 71 space
groups lead to purely Abelian flavor symmetries and we find no flavor in three cases.

In most cases, arriving at 4D models that reproduce properties of particle physics requires 
additional elements (such as D-branes, orientifols, Wilson lines, etc.) that break the flavor
symmetries we have classified to their subgroups.
In the heterotic strings compactified on Abelian toroidal orbifolds, their gauge 
embeddings admit the inclusion of Wilson lines, whose different values, restricted by modular invariance,
lead to a variety of effective field theories.

We have performed the widest known search of MSSM-like Abelian toroidal orbifolds of the
\E8\x\E8 heterotic string and found more than 121,000 promising models with different properties         
arising from orbifolds defined by the 64 space groups that yield non-Abelian flavor symmetries in 
our classification. Almost 115,000 models arise from different geometries of \Z{N}\x\Z{M} orbifolds,     
but only 422 arise from orbifolds with roto-translations, disfavoring this class of models               
for phenomenology. We show a summary of these results in tables~\ref{tab:ZNclassification} 
and~\ref{tab:ZNxZMclassification}. These models represent as many as nine times more models              
than those found in the literature.

Assuming that these models can describe some generic properties of the region of the string landscape
where the stringy ultraviolet completion of the SM resides, we have then studied the flavor properties
of these models. We have found that about 75.4\% of them exhibit flavor symmetries that are products             
of powers of $D_4$ and cyclic groups, whereas only about 1.2\% contain $\Delta(54)$. The remaining               
models are furnished with purely Abelian flavor symmetries, whose origin are the rules that dictate
how string states couple after compactification (see eqs.~\eqref{eq:pointgroupSel} and~\eqref{eq:transSel}). 
These results are summarized in the appendix~\ref{appendix},
and all model definitions are provided in our website~\cite{WebTables:2018xx}, where our promising models
are classified according to their space group, number of Wilson lines and flavor symmetry.

Two observations of these results are in order. Since models with $D_4$ flavor distinguish the 
third generation from the other two while models with $\Delta(54)$ frequently assign equal properties to all 
three generations, one can argue that our findings disfavor statistically the second scenario. 
Nonetheless, one may also be interested in studying the properties of the 
almost 1,500 models with $\Delta(54)$ as flavor structure.                                                       

Our second observation concerns the models with purely Abelian symmetries. It is somewhat surprising 
that almost one third of
our promising models have such flavor symmetries. Further, we must point out that we have not
performed a search of MSSM-like constructions for the 71 orbifold space groups that led to Abelian 
symmetries in the absence of WLs, according to table~\ref{tab:flavorsNoWLs}. We have only explored \Z6\x\Z6
and found almost 3,700 models endowed with a \Z6\x\Z6 flavor symmetry. Extrapolating this number of           
models, it is conceivable that most promising 
models arising from orbifold compactifications exhibit Abelian flavor symmetries. Therefore, we
consider necessary to further investigate the phenomenological implications of models whose flavor
structure coincide with the cyclic symmetry groups we find. We shall pursue this goal elsewhere.

On the other hand, although restricting ourselves to the massless sector of our models is
reasonable, the massive sectors of the compactification can also influence physics at low energies.
Thus, beside further research on the phenomenology of the models we have found and possible extensions
to string compactifications without supersymmetry, one should study how our findings are altered 
when massive strings or, in other words, all space group elements are considered. This is 
matter of ongoing research~\cite{Ramos-Sanchez:2018edc}.

Additionally, it is well-known that target-space modular symmetries act non-trivially on quarks 
and leptons arising from orbifold compactifications~\cite{Lauer:1989ax,Lauer:1990tm,Ferrara:1989qb,Bailin:1993ri}.
It has been recently emphasized the key role that these symmetries may play in flavor 
phenomenology~\cite{Feruglio:2017spp, Penedo:2018nmg, Kobayashi:2018rad, Kobayashi:2018scp, Kobayashi:2018vbk}. 
We find interesting to study systematically these symmetries in heterotic orbifolds, as shall be
done elsewhere.

Finally, one can extend the study of the stringy landscape of flavor physics by exploring 
non-Abelian orbifolds. Since the space group selection rule would most likely lead directly to non-Abelian symmetries, 
instead of only products of cyclic groups, one could expect a richer flavor structure. Following recent progress
on the understanding of these constructions~\cite{Fischer:2013qza}, we can pursue this
enterprise now.

\section*{Acknowledgments}
It is a pleasure to thank Patrick K.S. Vaudrevange and Hans-Peter Nilles for useful discussions.
SR-S is grateful to the Bethe Center for Theoretical Physics and the Mainz Institute for Theoretical
Physics for the hospitality during the final stage of this work.
This work was partly supported by DGAPA-PAPIIT grant IN100217 and CONACyT grants F-252167
and 278017.

\begin{appendix}
\begin{landscape}

\LTcapwidth=0.9\linewidth

\section{Flavor symmetries in models with Wilson Lines} 
\label{appendix}

{\footnotesize
\noindent
\begin{longtable}{|l|l|c|p{4cm}p{3.8cm}p{2.5cm}p{2.5cm}|c|}
 \caption{\label{tab:ZNflavorviable}Flavor symmetries in \Z{N} 6D orbifolds with WLs. The orbifold labels correspond to the labels of the
 associated space group, according to ref.~\cite{Fischer:2012qj}. For each space group yielding non-Abelian flavor
 symmetries in the absence of WLs (see table~\ref{tab:flavorsNoWLs}), we show all possible breakings for non-vanishing WLs. There are cases where the same 
 space group geometry and number of WLs lead to different flavor symmetries; these symmetries are stacked in different rows. Under each flavor symmetry, 
 we show the number of phenomenologically viable heterotic models obtained from our fairly exhaustive search of models (see section~\ref{sec:class}).}\\
\hline
\multicolumn{2}{|c|}{\multirow{2}{*}{Orbifold}}
                        &Max. \# of   & \multicolumn{4}{c|}{Flavor symmetry  with $\ell$ non-vanishing WLs}&\multirow{2}{*}{Total}\\
\multicolumn{2}{|c|}{}  &possible WLs & $\ell=$  1    &  2    &  3   & 4                                   &\\
\cline{4-7}                
\hline
\hline
\endfirsthead
\hline
\multicolumn{2}{|c|}{\multirow{2}{*}{Orbifold}} 
                        &Max. \# of   & \multicolumn{4}{c|}{Flavor symmetry with $\ell$ non-vanishing WLs}&\multirow{2}{*}{Total}\\
\multicolumn{2}{|c|}{}  &possible WLs & $\ell=$  1    &  2    &  3   & 4                                  &\\   
\hline
\hline
\endhead
\hline
\endfoot
\endlastfoot
\Z3     & (1,1)         & 3           & $\Delta(54)^2 $                       & $\Delta(54)\x\Z3^2$         & $\Z3^4$                && \multirow{2}{*}{0}\\
&&                                    &    0                                  &     0                       &     0                  &&\\ 
\hline
\Z4     & (1,1)         & 4           & $(D_4^3\x\Z4)/\Z2$                    & $D_4^2\x\Z4$                & $D_4\x\Z4\x\Z2^2$      & $\Z4\x\Z2^2$ ${}^{{}^{{}^{}}}$ & \multirow{2}{*}{0}\\
&&                                    &    0                                  &     0                       &     0                  &    0           &\\
\cline{2-8}
        & (2,1)         & 3           & $(S_2 \x S_2) \ltimes(\Z4^2 \x\Z2^2)$ &$S_2 \ltimes(\Z4^2 \x\Z2^2)$ & $\Z4^2 \x\Z2^2$ ${}^{{}^{{}^{{}^{}}}}$ && \multirow{4}{*}{149}\\
&&                                    &    0                                  &     0                       &   149                  &&\\
&&                                    & $(S_4 \x S_2) \ltimes(\Z4^3 \x\Z2^3)$ &$S_4\ltimes(\Z4^3 \x\Z2^3)$  &                        &&\\
&&                                    &    0                                  &     0                       &                        &&\\
\cline{2-8}
        & (3,1)         & 2           & $S_4\ltimes(\Z4^4\x \Z2)$             & $\Z4^3$                     & ${}^{{}^{{}^{{}^{}}}}$ && \multirow{2}{*}{27}\\
&&                                    &    0                                  &    27                       &                        &&\\
&&&&&&&\\
\hline
$\Z6$-I & (1,1)         & 1           & $\Z3\x\Z3$                            & ${}^{{}^{{}^{{}^{}}}}$      &                        && \multirow{2}{*}{30}\\
&&                                    &   30                                  &                             &                        &&\\
\cline{2-8}   
        & (2,1)         & 1           & $\Z6 \x \Z3$                          & ${}^{{}^{{}^{{}^{}}}}$      &                        && \multirow{2}{*}{30}\\
&&                                    &   30                                  &                             &                        &&\\  
\hline
$\Z6$-II& (1,1)         & 3           & $[(D_4 \x D_4) /\Z2]\x \Z3^2$         & $D_4 \x \Z2 \x \Z3^2$       & $\Z6\x\Z3\x\Z2^2$      & ${}^{{}^{{}^{{}^{}}}}$ &\multirow{4}{*}{363}\\
&&                                    &    0                                  &   337                       &    26                  &                        &\\
&&                                    & $\Delta(54)\x D_4 \x \Z2$             & $\Delta(54)\x \Z2^3$        &                        &&\\
&&                                    &    0                                  &     0                       &                        &&\\
\cline{2-8}
        & (2,1)         & 3           & $\Z6\x\Z3\x[(D_4\x D_4)/\Z2^2]$       & $D_4\x\Z6\x\Z3$             & $\Z6\x\Z3\x\Z2^{2}$    & ${}^{{}^{{}^{{}^{}}}}$ &\multirow{4}{*}{349}\\
&&                                    &    0                                  &   335                       &    14                  &&\\
&&                                    & $[(\Delta(54)\x \Z6)/\Z3] \x D_4$     & $[(\Delta(54)\x\Z6)/\Z3]\x\Z2^2$&                    &&\\
&&                                    &    0                                  &     0                       &                        &&\\
\cline{2-8}
        & (3,1)         & 3           & $\Z6\x \Z3 \x [(D_4 \x D_4)/\Z2^2]$   & $D_4\x\Z6\x\Z3$             & $\Z6\x \Z3 \x \Z2^{2}$ & ${}^{{}^{{}^{{}^{}}}}$ &\multirow{4}{*}{353}\\
&&                                    &    0                                  &   333                       &    18                  &&\\
&&                                    & $[(\Delta(54)\x \Z6)/\Z3] \x D_4$     & $[(\Delta(54)\x\Z6)/\Z3]\x\Z2^2$&                    &&\\
&&                                    &    0                                  &     2                       &                        &&\\
\cline{2-8}      
        & (4,1)         & 2           & $[(\Delta(54)\x \Z6)/\Z3] \x \Z2 $    & $\Z6 \x \Z3\x \Z2$          & ${}^{{}^{{}^{{}^{}}}}$ && \multirow{4}{*}{356}\\
&&                                    &    0                                  &    44                       &                        &&\\
&&                                    & $[D_4/\Z2]\x \Z6\x \Z3$               &                             &                        &&\\
&&                                    &  312                                  &                             &                        &&\\
\hline
\Z7     & (1,1)         & 1           & $\Z7^2$                               & ${}^{{}^{{}^{{}^{}}}}$      &&&\multirow{2}{*}{1} \\
&&                                    &    1                                  &                             &&&\\
\hline
$\Z8$-I & (1,1)         & 2           & $D_4\x\Z8$                            & $\Z8\x\Z2^2$                & ${}^{{}^{{}^{{}^{}}}}$ && \multirow{2}{*}{268}\\
&&                                    &   38                                  &   230                       &&&\\  
\cline{2-8}
        & (2,1)         & 2           & $D_4\x\Z8$                            & $\Z8\x\Z2^2$                & ${}^{{}^{{}^{{}^{}}}}$ && \multirow{2}{*}{246}\\
&&                                    &   41                                  &   205                       &&&\\    
\cline{2-8}        
        & (3,1)         & 1           & $\Z8\x\Z4$                            & ${}^{{}^{{}^{{}^{}}}}$      &&&\multirow{2}{*}{389}\\
&&                                    &  389                                  &                             &&&\\
\hline
\Z8-II  & (1,1)         & 3           & $(D_4\x D_4\x\Z8)/\Z2$                & $D_4\x\Z8\x\Z2$             & $\Z8\x\Z2^3$           & ${}^{{}^{{}^{{}^{}}}}$ &\multirow{2}{*}{2,023}\\
&&                                    &   21                                  &   398                       & 1,604                  &                        &\\
\cline{2-8} 
        & (2,1)         & 2           & $D_4 \x \Z8$                          &  $\Z8\x \Z2^2$              & ${}^{{}^{{}^{{}^{}}}}$ &                        & \multirow{2}{*}{505}\\
&&                                    &  231                                  &   274                       &                        &                        &\\
 
\hline
\Z{12}-I & (1,1)        & 1           & $\Z3\x \Z3$                           &&&& \multirow{2}{*}{556}\\
&&                                    &  556                                  &&&&\\
\cline{2-8}     
         & (2,1)        & 1           & $\Z{12}\x\Z3 $                        &&&&\multirow{2}{*}{555}\\
&&                                    &  555                                  &&&&\\
\hline
\Z{12}-II & (1,1)       & 2           & $D_4 \x \Z2$                          & $\Z2^3$                      & ${}^{{}^{{}^{{}^{}}}}$ &&\multirow{2}{*}{363}\\
&&                                    &    84                                 &   279                        &                        &&\\
\hline
\end{longtable}}

\newpage
\LTcapwidth=\linewidth
 {\footnotesize
 \begin{longtable}{|c|c|c|c|p{2.3cm}p{2.1cm}p{2cm}p{1.7cm}p{1.5cm}p{1cm}|c|} 
 \caption{\label{tab:ZNxZMflavorviable}Flavor symmetries in \Z{N}\x\Z{M} orbifolds with WLs. The orbifold labels correspond to the labels of the
 associated space group, according to ref.~\cite{Fischer:2012qj}. For each space group yielding non-Abelian flavor
 symmetries in the absence of WLs (see table~\ref{tab:flavorsNoWLs}), we show all possible breakings for non-vanishing WLs. There are cases where the same 
 space group geometry and number of WLs lead to different flavor symmetries; these symmetries are stacked in different rows. Under each flavor symmetry, 
 we show the number of phenomenologically viable models obtained from our fairly exhaustive search of models (see section~\ref{sec:class}).
 Since some WLs do not break the flavor group, we give in the fourth column the maximal number of WLs that affect the flavor symmetries.
 For \Z3\x\Z6 and \Z4\x\Z4  we also count in the total the promising models arising without WLs.}\\
 \hline
\multicolumn{2}{|c|}{\multirow{3}{*}{Orbifold}}& Max. \# of& Max. \# of & \multicolumn{6}{c|}{Flavor symmetry with $\ell$ non-vanishing WLs}&       \\
\multicolumn{2}{|c|}{}           & possible                & WLs affecting the &  $\ell=$ 1 &  2  &  3   &  4  &   5  &  6                  & Total \\
\multicolumn{2}{|c|}{}           & WLs                     & flavor symmetry   &            &     &      &     &      &                     &       \\
\hline
\hline
\endfirsthead
\hline
\multicolumn{2}{|c|}{\multirow{3}{*}{Orbifold}}& Max. \# of& Max. \# of & \multicolumn{6}{c|}{Flavor symmetry with $\ell$ non-vanishing WLs}&       \\
\multicolumn{2}{|c|}{}           & possible                 &WLs affecting the & $\ell=$ 1  &  2     &  3    &  4   &  5   &  6             & Total \\ 
\multicolumn{2}{|c|}{}           & WLs                      &flavor symmetry   &            &        &       &      &      &                &       \\
\hline
\hline
\endhead
 \hline
 \endfoot
 \endlastfoot
$\Z2\x \Z2$ & (1,1)    &\multirow{1}{*}{6}&\multirow{1}{*}{6}& $D_4^5 / \Z2^2$      & $D_4^4$             & $D_4^3\x\Z2^2$    & $D_4^2\x\Z2^4$         & $D_4\x\Z2^6$           & $\Z2^8$ ${}^{{}^{{}^{{}^{}}}}$ & \multirow{2}{*}{205}\\
&&&                                                          &       0              &       0             &      0            &      52                &  152                   & 1                      &\\ 
\cline{2-11}
&\multirow{1}{*}{(1,3)}&\multirow{1}{*}{4}&\multirow{1}{*}{2}& $D_4^2$              & $D_4 \x \Z2^2$      & $D_4 \x \Z2^2$    & $D_4 \x \Z2^2$         & ${}^{{}^{{}^{{}^{}}}}$ && \multirow{6}{*}{0} \\
&&&                                                          &       0              &       0             &      0            &       0                &                        &&\\
&&&                                                          & $D_4^3/\Z2^2$        & $D_4^2$             & $D_4^2$           &                        &                        &&\\  
&&&                                                          &       0              &       0             &      0            &                        &                        &&\\
&&&                                                          &                      & $D_4^3/\Z2^2$       &                   &                        &                        &&\\  
&&&                                                          &                      &       0             &                   &                        &                        &&\\
\cline{2-11}
&\multirow{1}{*}{(2,1)}&\multirow{1}{*}{5}&\multirow{1}{*}{5}& $D_4^4/ \Z2$         & $D_4^3 \x \Z2$      & $D_4^2\x\Z2^3$    & $D_4 \x \Z2^5$         & $\Z2^7$                & ${}^{{}^{{}^{{}^{}}}}$ &\multirow{2}{*}{369} \\
&&&                                                          &       0              &       0             &     14            &     342                &   13                   &                        &\\
 \cline{2-11}
&\multirow{1}{*}{(2,3)}&\multirow{1}{*}{3}&\multirow{1}{*}{2}& $D_4\x\Z2^2$         & $\Z2^4$             & $\Z2^4$           & ${}^{{}^{{}^{{}^{}}}}$ &&&\multirow{6}{*}{0} \\
&&&                                                          &       0              &       0             &      0            &                        &&&\\
&&&                                                          & $D_4^2$              & $D_4\x\Z2^2$        &                   &                        &&&\\
&&&                                                          &       0              &       0             &                   &                        &&&\\
&&&                                                          & $D_4^3/\Z2^2$        & $D_4^2$             &                   &                        &&&\\
&&&                                                          &       0              &       0             &                   &                        &&&\\
\cline{2-11}
&\multirow{1}{*}{(2,5)}&\multirow{1}{*}{3}&\multirow{1}{*}{1}& $D_4\x \Z2$          & $D_4\x \Z2$         & $D_4\x \Z2$       & ${}^{{}^{{}^{{}^{}}}}$ &&&\multirow{4}{*}{0}\\
&&&                                                          &       0              &       0             &      0            &                        &&&\\
&&&                                                          & $D_4^2 /\Z2$         & $D_4^2 /\Z2$        &                   &                        &&&\\
&&&                                                          &       0              &       0             &                   &                        &&&\\
\cline{2-11}
&\multirow{1}{*}{(3,1)}&\multirow{1}{*}{5}&\multirow{1}{*}{5}& $D_4^3$              & $D_4^2\x\Z2^2$      & $D_4 \x \Z2^4$    & $\Z2^6$                & $\Z2^6$                & ${}^{{}^{{}^{{}^{}}}}$ &\multirow{4}{*}{444}\\
&&&                                                          &       0              &       0             &     40            &       8                &    4                   &                        &\\
&&&                                                          &                      & $D_4^3$             & $D_4^2 \x\Z2^2$   & $D_4 \x\Z2^4$          &                        &                        &\\ 
&&&                                                          &                      &       0             &      0            &     392                &                                                          &&\\
\cline{2-11}
&\multirow{1}{*}{(3,3)}&\multirow{1}{*}{3}&\multirow{1}{*}{1}& $D_4\x \Z2$          & $D_4\x \Z2$         & $D_4\x \Z2$       & ${}^{{}^{{}^{{}^{}}}}$ &&&\multirow{4}{*}{0}\\
&&&                                                          &       0              &       0             &      0            &                        &&&\\
&&&                                                          & $D_4^2 /\Z2$         & $D_4^2 /\Z2$        &                   &                        &&&\\
&&&                                                          &       0              &       0             &                   &                        &&&\\
\cline{2-11}
&\multirow{1}{*}{(4,1)}&\multirow{1}{*}{4}&\multirow{1}{*}{2}& $D_4^3$              & $D_4^2\x \Z2^2$     & $D_4^2\x \Z2^2$   & $D_4^2\x\Z2^2$         & ${}^{{}^{{}^{{}^{}}}}$ &&\multirow{6}{*}{0}\\
&&&                                                          &       0              &       0             &      0            &       0                &                        &&\\
&&&                                                          & $D_4/\Z2^2$          & $D_4^3$             & $D_4^3$           &                        &                        &&\\
&&&                                                          &       0              &       0             &      0            &                        &                        &&\\
&&&                                                          &                      & $D_4/\Z2^2$         &                   &                        &                        &&\\
&&&                                                          &                      &       0             &                   &                        &                        &&\\
\cline{2-11}
&\multirow{1}{*}{(5,1)}&\multirow{1}{*}{4}&\multirow{1}{*}{4}& $D_4^3$              & $D_4^2\x \Z2^2$     & $D_4\x \Z2^4$     & $\Z2^6$                & ${}^{{}^{{}^{{}^{}}}}$ &&\multirow{2}{*}{42}\\
&&&                                                          &       0              &       0             &     40            &       2                &                        &&\\
\cline{2-11}
&\multirow{1}{*}{(5,4)}&\multirow{1}{*}{2}&\multirow{1}{*}{1}& $D_4\x\Z2$           & $D_4\x\Z2$          &                   &                        &&&\multirow{4}{*}{0} \\
&&&                                                          &       0              &       0             &                   &                        &&&\\
&&&                                                          & $D_4^2/ \Z2$         &                     &                   &                        &&&\\
&&&                                                          &       0              &                     &                   &                        &&&\\
\cline{2-11}
&\multirow{1}{*}{(6,1)}&\multirow{1}{*}{4}&\multirow{1}{*}{2}& $D_4^3 $             & $D_4^2\x\Z2^2$      & $D_4^2\x\Z2^2$    & $D_4^2\x\Z2^2$         & ${}^{{}^{{}^{{}^{}}}}$ &&\multirow{6}{*}{401} \\
&&&                                                          &       0              &       0             &     57            &    344                 &                        &&\\
&&&                                                          & $D_4^2 /\Z2^2$       & $ D_4^3 $           & $D_4^3$           &                        &                        &&\\
&&&                                                          &       0              &       0             &      0            &                        &                        &&\\
&&&                                                          &                      & $D_4^4 /\Z2^2$      &                   &                        &                        && \\
&&&                                                          &                      &       0             &                   &                        &                        &&\\

\cline{2-11}
&\multirow{1}{*}{(6,3)}&\multirow{1}{*}{2}&\multirow{1}{*}{0}& $D_4$                & $D_4$               &                   &                        &&&\multirow{2}{*}{0}\\
&&&                                                          &       0              &       0             &                   &                        &&& \\
\cline{2-11}
&\multirow{1}{*}{(7,1)}&\multirow{1}{*}{4}&\multirow{1}{*}{3}& $D_4 ^2\x \Z2$       & $D_4 \x\Z2^3$       & $D_4\x \Z2^3$     & $D_4\x\Z2^3$           & ${}^{{}^{{}^{{}^{}}}}$ &&\multirow{4}{*}{76}  \\
&&&                                                          &       0              &       0             &     55            &     21                 &                        &&\\
&&&                                                          & $D_4 ^3/\Z2$         & $D_4^2\x\Z2$        & $D_4^2\x \Z2$     &                        &                        &&\\
&&&                                                          &       0              &       0             &      0            &                        &                        &&\\
\cline{2-11}  
&\multirow{1}{*}{(8,1)}&\multirow{1}{*}{4}&\multirow{1}{*}{4}& $D_4 \x \Z2^2 $      & $\Z2^4$             & $\Z2^4$           & $\Z2^4$                & ${}^{{}^{{}^{{}^{}}}}$ &&\multirow{4}{*}{25}  \\
&&&                                                          &       0              &       0             &      0            &     25                 &                        &&\\
&&&                                                          &                      & $D_4 \x \Z2^2 $     &                   &                        &                        &&\\
&&&                                                          &                      &       0             &                   &                        &                        &&\\
\cline{2-11}
&\multirow{1}{*}{(9,1)}&\multirow{1}{*}{3}&\multirow{1}{*}{2}& $D_4^2 \x\Z2$        & $D_4 \x\Z2^3$       & $D_4 \x\Z2^3$     & ${}^{{}^{{}^{{}^{}}}}$ &&&\multirow{2}{*}{27}\\
&&&                                                          &       0              &       2             &     25            &                        &&&\\
&&&                                                          & $D_4^3/ \Z2$         & $D_4^2 \x\Z2$       &                   &                        &&&\\
&&&                                                          &       0              &       0             &                   &                        &&&\\
\cline{2-11}
&\multirow{1}{*}{(10,1)}&\multirow{1}{*}{3}&\multirow{1}{*}{3}& $D_4\x \Z2^2 $      & $\Z2^4 $            & $\Z2^4$           & ${}^{{}^{{}^{{}^{}}}}$ &&& \multirow{4}{*}{21}\\
&&&                                                           &      0              &       2             &     19            &                        &&&\\
&&&                                                           &                     & $D_4\x\Z2^2$        &                   &                        &&&\\
&&&                                                           &                     &       0             &                   &                        &&&\\
\cline{2-11}   
&\multirow{1}{*}{(11,1)}&\multirow{1}{*}{3}&\multirow{1}{*}{0}& $D_4^3/\Z2 $        & $D_4^3/\Z2 $        & $D_4^3/\Z2$       & ${}^{{}^{{}^{{}^{}}}}$ &&&\multirow{2}{*}{0}\\
&&&                                                           &      0              &       0             &       0           &                        &&&                  \\
\cline{2-11}  
&\multirow{1}{*}{(12,1)}&\multirow{1}{*}{2}&\multirow{1}{*}{2}& $D_4\x \Z2^2$       & $\Z2^4$             &                   & ${}^{{}^{{}^{{}^{}}}}$ &&& \multirow{2}{*}{3}\\
&&&                                                           &      0              &       3             &                   &                        &&& \\
\hline
$\Z2\x \Z4$ & (1,1)    &\multirow{1}{*}{4}&\multirow{1}{*}{4}& $(D_4^3 \x\Z4)/\Z2$  &  $D_4^2\x\Z4\x\Z2$  & $D_4\x\Z4\x\Z2^3$ & $\Z4\x\Z2^5$           & ${}^{{}^{{}^{{}^{}}}}$ && \multirow{2}{*}{10,580} \\
&&&                                                          &      26              &   1,463             &   8,637           &    454                 &                        &&\\
\cline{2-11} 
&\multirow{1}{*}{(1,6)}&\multirow{1}{*}{2}&\multirow{1}{*}{2}& $D_4^2\x\Z4$         & $D_4\x\Z4\x \Z2^2$  &                   & ${}^{{}^{{}^{{}^{}}}}$ &&&\multirow{2}{*}{86}\\
&&&                                                          &      21              &      65             &                   &                        &&&\\
\cline{2-11}  
&\multirow{1}{*}{(2,1)}&\multirow{1}{*}{4}&\multirow{1}{*}{4}& $(D_4^3\x\Z4)/\Z2$   & $D_4^2\x\Z4\x\Z2$   & $D_4\x\Z4\x\Z2^3$ & $\Z4\x\Z2^5$           & ${}^{{}^{{}^{{}^{}}}}$ &&\multirow{2}{*}{6,158}  \\
&&&                                                          &      81              &   1,131             &   4,686           &    260                 &                        &&\\
\cline{2-11}
&\multirow{1}{*}{(2,4)}&\multirow{1}{*}{2}&\multirow{1}{*}{2}& $D_4\x \Z4\x\Z2$     & $\Z4\x\Z2^3$        &                   & ${}^{{}^{{}^{{}^{}}}}$ &&&\multirow{2}{*}{328} \\
&&&                                                          &      47              &     281             &                   &                        &&&\\
\cline{2-11} 
&\multirow{1}{*}{(3,1)}&\multirow{1}{*}{3}&\multirow{1}{*}{2}& $D_4^2 \x\Z4$        & $D_4\x\Z4 \x \Z2^2$ & $D_4\x\Z4\x\Z2^2$ & ${}^{{}^{{}^{{}^{}}}}$ &&&\multirow{4}{*}{22,305}\\
&&&                                                          &      27              &   1,109             &  18,440           &                        &&&  \\
&&&                                                          & $(D_4^3\x\Z4)/\Z2^2$ & $D_4^2\x\Z4$        &                   &                        &&&  \\
&&&                                                          &      76              &   2,653             &                   &                        &&&  \\
\cline{2-11}
&\multirow{1}{*}{(4,1)}&\multirow{1}{*}{3}&\multirow{1}{*}{3}& $D_4 \x \Z4\x \Z2$   & $\Z4\x \Z2^3$       & $\Z4\x\Z2^3$      & ${}^{{}^{{}^{{}^{}}}}$ &&&\multirow{4}{*}{4,519}\\
&&&                                                          &      33              &   1,133             &   2,911           &                        &&& \\
&&&                                                          &                      & $D_4\x\Z4\x\Z2$     &                   &                        &&& \\
&&&                                                          &                      &     442             &                   &                        &&& \\
\cline{2-11} 
&\multirow{1}{*}{(5,1)}&\multirow{1}{*}{3}&\multirow{1}{*}{2}& $D_4^2\x \Z4$        & $D_4\x\Z4 \x \Z2^2$ & $D_4\x\Z4\x\Z2^2$ & ${}^{{}^{{}^{{}^{}}}}$ &&&\multirow{2}{*}{2,116}\\
&&&                                                          &      18              &      45             &   1,311           &                        &&& \\
&&&                                                          & $(D_4^3\x \Z4)/\Z2^2$& $D_4^2\x\Z4$        &                   &                        &&& \\
&&&                                                          &      45              &     697             &                   &                        &&& \\
\cline{2-11} 
&\multirow{1}{*}{(6,1)}&\multirow{1}{*}{3}&\multirow{1}{*}{2}& $D_4 \x\Z4 \x \Z2^2$ & $\Z4 \x \Z2^4$      & $\Z4\x\Z2^4$      & ${}^{{}^{{}^{{}^{}}}}$ &&&\multirow{6}{*}{3,246}  \\
&&&                                                          &      18              &     511             &   1,814           &                        &&& \\
&&&                                                          & $D_4^2\x\Z4 $        & $D_4\x \Z4\x \Z2^2$ &                   &                        &&& \\
&&&                                                          &       3              &     295             &                   &                        &&& \\
&&&                                                          & $(D_4^3\x\Z4)/\Z2$   & $D_4^2 \x \Z4$      &                   &                        &&& \\
&&&                                                          &      37              &     568             &                   &                        &&& \\
\cline{2-11} 
&\multirow{1}{*}{(7,1)}&\multirow{1}{*}{3}&\multirow{1}{*}{3}& $D_4\x\Z4\x\Z2$      & $\Z4\x\Z2^3$        & $\Z4\x\Z2^3$      & ${}^{{}^{{}^{{}^{}}}}$ &&&\multirow{4}{*}{2,667} \\
&&&                                                          &      64              &     729             &   1,481           &                        &&& \\
&&&                                                          &                      & $D_4\x\Z4\x\Z2$     &                   &                        &&& \\
&&&                                                          &                      &     393             &                   &                        &&& \\
\cline{2-11} 
&\multirow{1}{*}{(8,1)}&\multirow{1}{*}{2}&\multirow{1}{*}{0}& $(D_4^2\x\Z4)/ \Z2$  & $(D_4^2\x\Z4)/ \Z2$ &                   & ${}^{{}^{{}^{{}^{}}}}$ &&&\multirow{2}{*}{911}  \\
&&&                                                          &      72              &     839             &                   &                        &&& \\
\cline{2-11} 
&\multirow{1}{*}{(9,1)}&\multirow{1}{*}{2}&\multirow{1}{*}{2}& $D_4 \x \Z4\x \Z2$   & $\Z4 \x \Z2^3$      &                   & ${}^{{}^{{}^{{}^{}}}}$ &&&\multirow{2}{*}{2,142}  \\
&&&                                                          &     522              &   1,620             &                   &                        &&& \\
\hline
$\Z2\x\Z6$-I & (1,1)   &\multirow{1}{*}{2}&\multirow{1}{*}{2}& $D_4\x\Z2\x\Z6$      & $\Z2^3\x\Z6$        &                   & ${}^{{}^{{}^{{}^{}}}}$ &&&\multirow{2}{*}{583}\\
&&&                                                          &     116              &     467             &                   &                        &&& \\
\cline{2-11}   
&\multirow{1}{*}{(2,1)}&\multirow{1}{*}{2}&\multirow{1}{*}{2}& $D_4\x\Z2\x\Z6$      & $\Z2^3\x\Z6$        &                   & ${}^{{}^{{}^{{}^{}}}}$ &&&\multirow{2}{*}{353}\\
&&&                                                          &      78              &     275             &                   &                        &&& \\
\hline
$\Z3\x \Z3$& (1,1)     &\multirow{1}{*}{3}&\multirow{1}{*}{3}& $\Delta(54)^2\x\Z3$  & $\Delta(54)\x\Z3^3$ & $\Z3^5$           & ${}^{{}^{{}^{{}^{}}}}$ &&&\multirow{2}{*}{1,108}\\
&&&                                                          &      81              &     987             &      40           &                        &&& \\
\cline{2-11}
&\multirow{1}{*}{(1,4)}&\multirow{1}{*}{1}&\multirow{1}{*}{1}& $\Z3^3$              &                     &                   & ${}^{{}^{{}^{{}^{}}}}$ &&&\multirow{2}{*}{8}\\
&&&                                                          &       8              &                     &                   &                        &&& \\
\cline{2-11}
&\multirow{1}{*}{(2,1)}&\multirow{1}{*}{2}&\multirow{1}{*}{2}& $\Delta(54)\x \Z3^2$ & $\Z3^4$             &                   & ${}^{{}^{{}^{{}^{}}}}$ &&&\multirow{2}{*}{1,952}\\
&&&                                                          &     239              &   1,713             &                   &                        &&& \\
\cline{2-11}
&\multirow{1}{*}{(3,1)}&\multirow{1}{*}{2}&\multirow{1}{*}{2}& $\Z3^3$              & $\Z3^3$             &                   & ${}^{{}^{{}^{{}^{}}}}$ &&&\multirow{2}{*}{6}\\
&&&                                                          &       0              &       6             &                   &                        &&& \\
\cline{2-11}
&\multirow{1}{*}{(4,1)}&\multirow{1}{*}{2}&\multirow{1}{*}{1}& $\Z3^4$              & $\Z3^4$             &                   & ${}^{{}^{{}^{{}^{}}}}$ &&&\multirow{4}{*}{215}\\
&&&                                                          &      22              &     105             &                   &                        &&& \\
&&&                                                          & $\Delta(54)^2$       &                     &                   &                        &&& \\
&&&                                                          &      88              &                     &                   &                        &&& \\
\hline
$\Z3\x \Z6$& (1,1)     &\multirow{1}{*}{1}&\multirow{1}{*}{1}& $\Z3^2 \x \Z6$       &                     &                   & ${}^{{}^{{}^{{}^{}}}}$ &&&\multirow{2}{*}{4,493} \\
&&&                                                          &   4,469              &                     &                   &                        &&& \\
\cline{2-11} 
&\multirow{1}{*}{(2,1)}&\multirow{1}{*}{1}&\multirow{1}{*}{1}& $\Z3^2 \x \Z6$       &                     &                   & ${}^{{}^{{}^{{}^{}}}}$ &&& \multirow{2}{*}{540} \\
&&&                                                          &     495              &                     &                   &                        &&& \\
\hline
$\Z4\x \Z4$& (1,1)     &\multirow{1}{*}{3}&\multirow{1}{*}{3}& $(D_4^2\x\Z4^2)/\Z2$ & $D_4 \x\Z4^2\x\Z2$  & $\Z4^2 \x \Z2^3$  & ${}^{{}^{{}^{{}^{}}}}$ &&&\multirow{2}{*}{28,649} \\
&&&                                                          &   2,442              &  24,693             &     1,509           &                        &&&\\
&&& &&&&&&&\\
\cline{2-11}
&\multirow{1}{*}{(2,1)}&\multirow{1}{*}{2}&\multirow{1}{*}{1}& $D_4 \x \Z4^2$       &  $D_4 \x \Z4^2$     & ${}^{{}^{{}^{{}^{}}}}$ &                   &&&\multirow{2}{*}{9,853} \\
&&&                                                          &     556              &   6,286             &                   &                        &&& \\
&&&                                                          & $(D_4^2\x\Z4^2)/\Z2^2$&                   &                   &                        &&& \\
&&&                                                          &   2,992              &                     &                   &                        &&& \\
\cline{2-11}
&\multirow{1}{*}{(3,1)}&\multirow{1}{*}{2}&\multirow{1}{*}{2}& $D_4 \x \Z4^2$       & $D_4 \x \Z4^2$      & ${}^{{}^{{}^{{}^{}}}}$ &                   &&&\multirow{2}{*}{5522}\\
&&&                                                          &     1,003              &   4,513           &                   &                        &&&\\
\cline{2-11}   
&\multirow{1}{*}{(4,1)}&\multirow{1}{*}{2}&\multirow{1}{*}{1}& $\Z4^2\x \Z2^2$      & $\Z4^2 \x \Z2^2$    & ${}^{{}^{{}^{{}^{}}}}$ &                   &&&\multirow{2}{*}{4,730} \\
&&&                                                          &     423              &      3,097          &                   &                        &&&\\
&&&                                                          & $(D_4^2\x\Z4^2)/\Z2^2$& ${}^{{}^{{}^{{}^{}}}}$&                &                        &&&\\
&&&                                                          &   1,204              &                     &                   &                        &&&\\
\hline
\end{longtable}}
\end{landscape}
\end{appendix}


\providecommand{\bysame}{\leavevmode\hbox to3em{\hrulefill}\thinspace}
\frenchspacing
\newcommand{\origttfamily}{}
\let\origttfamily=\ttfamily
\renewcommand{\ttfamily}{\origttfamily \hyphenchar\font=`\-}

\end{document}